\documentclass[twocolumn]{aastex63}

\usepackage{graphicx}
\usepackage{xcolor}
\let\tablenum\relax
\usepackage{siunitx}
\usepackage[utf8]{inputenc} 
\usepackage[T5,T1]{fontenc}
\usepackage{mathtools} 
\usepackage[version=4]{mhchem} 
\usepackage[style=american]{csquotes} 
\usepackage{lineno}
\usepackage{longtable}
\usepackage{amsmath}  
\usepackage{amssymb}  
\usepackage{paralist} 
\usepackage{etoolbox} 
\usepackage[]{natbib}
\usepackage{hyperref}
\usepackage{placeins}
\usepackage[toc]{appendix}

\newcommand*\linenomathpatch[1]{%
  \cspreto{#1}{\linenomath}%
  \cspreto{#1*}{\linenomath}%
  \csappto{end#1}{\endlinenomath}%
  \csappto{end#1*}{\endlinenomath}%
}
\linenomathpatch{equation}
\linenomathpatch{gather}
\linenomathpatch{multline}
\linenomathpatch{align}
\linenomathpatch{alignat}
\linenomathpatch{flalign}

\newcommand{\nubar}{\overline{\nu}}

\received{receipt date}
\revised{revision date}
\accepted{acceptance date}
\published{published date}
\submitjournal{Astrophysical Journal}
\setwatermarkfontsize{64pt}
\shorttitle{Energy Spectrum of Astrophysical Muon Neutrinos}
\shortauthors{IceCube}

\begin{document}
\title{Improved Characterization of the Astrophysical Muon-Neutrino Flux with 9.5 Years of IceCube Data}

\correspondingauthor{The IceCube Collaboration}
\email{analysis@icecube.wisc.edu}
\affiliation{III. Physikalisches Institut, RWTH Aachen University, D-52056 Aachen, Germany}
\affiliation{Department of Physics, University of Adelaide, Adelaide, 5005, Australia}
\affiliation{Dept. of Physics and Astronomy, University of Alaska Anchorage, 3211 Providence Dr., Anchorage, AK 99508, USA}
\affiliation{Dept. of Physics, University of Texas at Arlington, 502 Yates St., Science Hall Rm 108, Box 19059, Arlington, TX 76019, USA}
\affiliation{CTSPS, Clark-Atlanta University, Atlanta, GA 30314, USA}
\affiliation{School of Physics and Center for Relativistic Astrophysics, Georgia Institute of Technology, Atlanta, GA 30332, USA}
\affiliation{Dept. of Physics, Southern University, Baton Rouge, LA 70813, USA}
\affiliation{Dept. of Physics, University of California, Berkeley, CA 94720, USA}
\affiliation{Lawrence Berkeley National Laboratory, Berkeley, CA 94720, USA}
\affiliation{Institut f{\"u}r Physik, Humboldt-Universit{\"a}t zu Berlin, D-12489 Berlin, Germany}
\affiliation{Fakult{\"a}t f{\"u}r Physik {\&} Astronomie, Ruhr-Universit{\"a}t Bochum, D-44780 Bochum, Germany}
\affiliation{Universit{\'e} Libre de Bruxelles, Science Faculty CP230, B-1050 Brussels, Belgium}
\affiliation{Vrije Universiteit Brussel (VUB), Dienst ELEM, B-1050 Brussels, Belgium}
\affiliation{Department of Physics and Laboratory for Particle Physics and Cosmology, Harvard University, Cambridge, MA 02138, USA}
\affiliation{Dept. of Physics, Massachusetts Institute of Technology, Cambridge, MA 02139, USA}
\affiliation{Dept. of Physics and Institute for Global Prominent Research, Chiba University, Chiba 263-8522, Japan}
\affiliation{Department of Physics, Loyola University Chicago, Chicago, IL 60660, USA}
\affiliation{Dept. of Physics and Astronomy, University of Canterbury, Private Bag 4800, Christchurch, New Zealand}
\affiliation{Dept. of Physics, University of Maryland, College Park, MD 20742, USA}
\affiliation{Dept. of Astronomy, Ohio State University, Columbus, OH 43210, USA}
\affiliation{Dept. of Physics and Center for Cosmology and Astro-Particle Physics, Ohio State University, Columbus, OH 43210, USA}
\affiliation{Niels Bohr Institute, University of Copenhagen, DK-2100 Copenhagen, Denmark}
\affiliation{Dept. of Physics, TU Dortmund University, D-44221 Dortmund, Germany}
\affiliation{Dept. of Physics and Astronomy, Michigan State University, East Lansing, MI 48824, USA}
\affiliation{Dept. of Physics, University of Alberta, Edmonton, Alberta, Canada T6G 2E1}
\affiliation{Erlangen Centre for Astroparticle Physics, Friedrich-Alexander-Universit{\"a}t Erlangen-N{\"u}rnberg, D-91058 Erlangen, Germany}
\affiliation{Physik-department, Technische Universit{\"a}t M{\"u}nchen, D-85748 Garching, Germany}
\affiliation{D{\'e}partement de physique nucl{\'e}aire et corpusculaire, Universit{\'e} de Gen{\`e}ve, CH-1211 Gen{\`e}ve, Switzerland}
\affiliation{Dept. of Physics and Astronomy, University of Gent, B-9000 Gent, Belgium}
\affiliation{Dept. of Physics and Astronomy, University of California, Irvine, CA 92697, USA}
\affiliation{Karlsruhe Institute of Technology, Institute for Astroparticle Physics, D-76021 Karlsruhe, Germany }
\affiliation{Karlsruhe Institute of Technology, Institute of Experimental Particle Physics, D-76021 Karlsruhe, Germany }
\affiliation{Dept. of Physics, Engineering Physics, and Astronomy, Queen's University, Kingston, ON K7L 3N6, Canada}
\affiliation{Dept. of Physics and Astronomy, University of Kansas, Lawrence, KS 66045, USA}
\affiliation{Department of Physics and Astronomy, UCLA, Los Angeles, CA 90095, USA}
\affiliation{Centre for Cosmology, Particle Physics and Phenomenology - CP3, Universit{\'e} catholique de Louvain, Louvain-la-Neuve, Belgium}
\affiliation{Department of Physics, Mercer University, Macon, GA 31207-0001, USA}
\affiliation{Dept. of Astronomy, University of Wisconsin{\textendash}Madison, Madison, WI 53706, USA}
\affiliation{Dept. of Physics and Wisconsin IceCube Particle Astrophysics Center, University of Wisconsin{\textendash}Madison, Madison, WI 53706, USA}
\affiliation{Institute of Physics, University of Mainz, Staudinger Weg 7, D-55099 Mainz, Germany}
\affiliation{Department of Physics, Marquette University, Milwaukee, WI, 53201, USA}
\affiliation{Institut f{\"u}r Kernphysik, Westf{\"a}lische Wilhelms-Universit{\"a}t M{\"u}nster, D-48149 M{\"u}nster, Germany}
\affiliation{Bartol Research Institute and Dept. of Physics and Astronomy, University of Delaware, Newark, DE 19716, USA}
\affiliation{Dept. of Physics, Yale University, New Haven, CT 06520, USA}
\affiliation{Dept. of Physics, University of Oxford, Parks Road, Oxford OX1 3PU, UK}
\affiliation{Dept. of Physics, Drexel University, 3141 Chestnut Street, Philadelphia, PA 19104, USA}
\affiliation{Physics Department, South Dakota School of Mines and Technology, Rapid City, SD 57701, USA}
\affiliation{Dept. of Physics, University of Wisconsin, River Falls, WI 54022, USA}
\affiliation{Dept. of Physics and Astronomy, University of Rochester, Rochester, NY 14627, USA}
\affiliation{Department of Physics and Astronomy, University of Utah, Salt Lake City, UT 84112, USA}
\affiliation{Oskar Klein Centre and Dept. of Physics, Stockholm University, SE-10691 Stockholm, Sweden}
\affiliation{Dept. of Physics and Astronomy, Stony Brook University, Stony Brook, NY 11794-3800, USA}
\affiliation{Dept. of Physics, Sungkyunkwan University, Suwon 16419, Korea}
\affiliation{Institute of Basic Science, Sungkyunkwan University, Suwon 16419, Korea}
\affiliation{Dept. of Physics and Astronomy, University of Alabama, Tuscaloosa, AL 35487, USA}
\affiliation{Dept. of Astronomy and Astrophysics, Pennsylvania State University, University Park, PA 16802, USA}
\affiliation{Dept. of Physics, Pennsylvania State University, University Park, PA 16802, USA}
\affiliation{Dept. of Physics and Astronomy, Uppsala University, Box 516, S-75120 Uppsala, Sweden}
\affiliation{Dept. of Physics, University of Wuppertal, D-42119 Wuppertal, Germany}
\affiliation{DESY, D-15738 Zeuthen, Germany}

\author[0000-0001-6141-4205]{R. Abbasi}
\affiliation{Department of Physics, Loyola University Chicago, Chicago, IL 60660, USA}

\author[0000-0001-8952-588X]{M. Ackermann}
\affiliation{DESY, D-15738 Zeuthen, Germany}

\author{J. Adams}
\affiliation{Dept. of Physics and Astronomy, University of Canterbury, Private Bag 4800, Christchurch, New Zealand}

\author[0000-0003-2252-9514]{J. A. Aguilar}
\affiliation{Universit{\'e} Libre de Bruxelles, Science Faculty CP230, B-1050 Brussels, Belgium}

\author[0000-0003-0709-5631]{M. Ahlers}
\affiliation{Niels Bohr Institute, University of Copenhagen, DK-2100 Copenhagen, Denmark}

\author{M. Ahrens}
\affiliation{Oskar Klein Centre and Dept. of Physics, Stockholm University, SE-10691 Stockholm, Sweden}

\author[0000-0002-9534-9189]{J.M. Alameddine}
\affiliation{Dept. of Physics, TU Dortmund University, D-44221 Dortmund, Germany}

\author{C. Alispach}
\affiliation{D{\'e}partement de physique nucl{\'e}aire et corpusculaire, Universit{\'e} de Gen{\`e}ve, CH-1211 Gen{\`e}ve, Switzerland}

\author{A. A. Alves Jr.}
\affiliation{Karlsruhe Institute of Technology, Institute for Astroparticle Physics, D-76021 Karlsruhe, Germany }

\author{N. M. Amin}
\affiliation{Bartol Research Institute and Dept. of Physics and Astronomy, University of Delaware, Newark, DE 19716, USA}

% ...

\date{\today}

\collaboration{379}{IceCube Collaboration}


\begin{abstract}
We present a measurement of the high-energy
astrophysical muon-neutrino flux with the IceCube Neutrino
Observatory. The measurement uses a high-purity selection of ~650k neutrino-induced muon tracks from the Northern celestial hemisphere, corresponding to 9.5 years of experimental data. With respect to previous publications, the measurement is improved by the increased size of the event sample and the extended model testing beyond simple power-law hypotheses. An updated treatment of systematic uncertainties and atmospheric background fluxes has been implemented based on recent models. The best-fit single power-law parameterization for the
astrophysical energy spectrum results in a normalization of $\phi_{\mathrm{@100TeV}}^{\nu_\mu+\bar{\nu}_\mu} = 1.44_{-0.26}^{+0.25} \times  10^{-18}\,\mathrm{GeV}^{-1}\mathrm{cm}^{-2}\mathrm{s}^{-1}\mathrm{sr}^{-1}$ and a spectral index $\gamma_{\mathrm{SPL}} = 2.37_{-0.09}^{+0.09}$, constrained
in the energy range from $15\,\mathrm{TeV}$ to $5\,\mathrm{PeV}$. The model tests include a single power law with a spectral cutoff at high energies, a log-parabola model, several source-class specific flux 
predictions from the literature and a model-independent spectral
unfolding. The data is well consistent with a single power law hypothesis, however, spectra with softening above one PeV are statistically more favorable at a two sigma level.
\end{abstract}

\section{Introduction}

The field of neutrino astronomy has gained momentum since the IceCube collaboration discovered a diffuse flux of astrophysical neutrinos in multiple detection channels~\citep{PhysRevLett.111.021103,aartsenObservationCharacterizationCosmic2016,aartsenCharacteristicsDiffuseAstrophysical2020}. Prominent examples of this continuous journey are the identification of a first joint source of high-energy gamma rays and neutrinos, TXS 0506+056~\citep{icecubeMultimessengerObservationsFlaring2018,aartsenNeutrinoEmissionDirection2018}, the increasing hints for the emission of high-energy neutrinos from the radio galaxy NGC1068~\citep{Aartsen:2019fau}, and detection of a particle shower at the Glashow resonance energy \citep{AartsenGlashowDetectionIcecube2021}.

Measuring the total observed flux strength and energy spectrum of high-energy astrophysical neutrinos complements direct searches for neutrino sources and is important to understand the processes behind the acceleration and propagation of high-energy cosmic rays. 

The origin of high-energy neutrinos has been predicted from non-thermal Galactic and extragalactic sources, as reviewed in  \cite{learnedHighEnergyNeutrinoAstrophysics2000, BeckerHighEnergyNuMultimessenenger2008, HalzenIceCubeInstrumentNuAstronomy2010} and \cite{HalzenOpeningWindowUniverseIceCube2018}. As the diffuse flux detected by IceCube is close to isotropically distributed and does not follow the Galactic plane, Galactic sources have been disfavoured, while still being discussed as a sub-dominant contribution, see \cite{BeckerClosingInOnOriginGalacticCR2020} for a summary. The prompt phase of Gamma Ray Bursts (GRBs) has also been excluded as a dominant neutrino source by dedicated IceCube analyses \citep{AbbasiAbsenceNuInGRB2012Nature, aartsenGRBlimit2017}. 

However, prominent possible sources of neutrinos exist, for example choked GRBs in dense environments  \citep{sennoChokedJetsLowLuminosity2016, biehlCosmicRayNeutrinoEmission2018}. A second, promising source of high-energetic extragalactic sources are active galactic nuclei (AGNs) \citep{muraseDiffuseNeutrinoIntensity2014, kimuraNeutrinoCosmicrayEmission2015, liuCanWindsDriven2018}, including BL-Lac objects \citep{tavecchioHighenergyCosmicNeutrinos2014, padovaniSimplifiedViewBlazars2015}. Tidal Disruption Events (TDEs) are a third promising source class of high-energy neutrinos and ultra-high energy cosmic rays \citep{FarrarTDEJetsUHECRSource2014, SennoNuFlaresXRayTDEs2017, LunardiniHENuTDEStars2017, DaiTDEproduceIceCubeNu2017, GuepinObserveNuFlaresCoincWExplosiveTransients2017, BiehlTDStarsAsCRandNuOrigin2018, SteinTDECoincHENu2021}. Here, particle acceleration could be driven by either a hidden jet, a hidden sub-relativistic wind, emission from a hot corona above the accretion disk, or radiatively inefficient accretion flows (see \cite{HayasakiNufromTDE2021} for a review). Starburst galaxies have also been considered as a possibly contributing source class \citep{LoebBackgroundNuFromStarburstGalaxies2006, KohtaStarFormingGalaxiesAsDiffuse2014}. These models have in common that they expect neutrinos to be produced when a power-law distributed population of cosmic rays interact with gas or photon fields in the source or its vicinity in order to produce pions and kaons, which in turn decay into neutrinos. These neutrinos would follow the same powerlaw as the initial cosmic-ray population, however, effects like an energy-dependent cross section, breaks, or spectral features in the accelerated cosmic-ray spectrum can lead to deviations from a pure power-law. 
In this paper, we present an improved measurement of the energy spectrum of astrophysical muon neutrinos, including models beyond the single power law and tests of a selection of the above-mentioned models. Also, we update the sample of up-going muon tracks ($\theta_{\mathrm{zenith}} > 85^\circ$) originating from the Northern celestial hemisphere~\citep{aartsenObservationCharacterizationCosmic2016}. 

\section{Data sample}
\label{sec:sample}

The IceCube Neutrino Observatory is a gigaton-scale Cherenkov detector embedded in the Antarctic ice at the geographical South Pole~\citep{aartsenIceCubeNeutrinoObservatory2017}. Its fundamental building blocks are 5160 digital optical modules (DOMs). These spherical detection units each include a $10\,''$ ($\SI{25.4}{cm}$) photomultiplier tube suited to detect weak light signals \citep{Abbasi:2010PMT} and read-out and digitization electronics \citep{abbasiIceCubeDataAcquisition2009}, and are positioned on 86 vertical cable strings which are arranged in a hexagonal grid. In this analysis, the data from 8 strings forming the denser instrumented region of the \textit{DeepCore} array \citep{AbbasiDeepCore2011} are excluded in favor of a more homogeneous detector geometry. Neutrinos are detected indirectly via the Cherenkov light emitted by charged relativistic secondary particles. These emerge from deep inelastic neutrino nucleon interactions, which occur in the instrumented and the surrounding ice or the bedrock below IceCube. The final detector configuration (IC86) was completed in December of 2010. Recording of data already occurred prior to finalization with partial configurations. Data from two such partial configurations, IC59 and IC79, are included in this analysis, with the number indicating the number of active strings.

The presented study is based on a sample of muon-neutrino induced muon tracks detected between May 20, 2009 and December 31, 2018. Muons can travel through the ice for hundreds or thousands of meters, producing \textit{track-like} signatures in the detector. The event selection is identical to the one presented in \cite{aartsenObservationCharacterizationCosmic2016}. It focuses on events with extended track-like signatures and filters out \textit{cascade-like} events which can for example arise from electron-neutrino induced electrons, which travel relatively short distances compared to the detector string spacing of $\SI{125}{m}$. By restricting the field of view of the detector to zenith angles $\theta_{\mathrm{zenith}} > 85^\circ$, the overwhelming background of atmospheric muons from cosmic-ray induced air showers is successfully suppressed (purity $>99.8\%$) as these muons are stopped in the Earth or the ice overburden before reaching IceCube. The direction of the selected muon-track events is obtained using the \textit{MPE} algorithm with tabulated ice properties~\citep{ahrensMuonTrackReconstruction2004,kaiStackedSearchesHighenergy2014} and the muon energy is reconstructed using the truncated energy (DOM's method) algorithm described in \cite{abbasiImprovedMethodMeasuring2013} which takes the stochastic energy losses of high-energy muons into account, resulting in an energy resolution of $0.22$ in $\log_{10}E_{\mu}$ \citep{abbasiImprovedMethodMeasuring2013}. The resulting proxy energy $E_{\mu, \mathrm{proxy}}$ is inherently only a lower limit on the true muon and primary neutrino energies, as the neutrino-nucleon interactions can occur outside the detection volume.

\begin{table*}
\centering
\begin{tabular}{lcccc}
			\hline
			{Data-Taking Season} &  {Zenith Range / deg} & {Effective Livetime$^{*}$ / years} & {Number of Events} & {Pass-2}\\
			\hline
			IC59               &  $90 - 180$ &  $0.9538$ & 21411 & No \\
			IC79 -- IC86-2018  &  $85 - 180$ & $8.181$ & 651377 & Yes \\
			Total		   &  				 & $9.135$ & 672788 &  \\
 			\hline
\end{tabular}
\caption{Summary of the event sample. See text for a description of the Pass-2 re-calibration campaign that was conducted for all data after May 2010.\label{tab:sample_properties}\\
$^{*}$The livetime for the season IC79 has been corrected by a factor $0.94$ to account for the lower expected trigger rate of the partial detector configuration.}
\end{table*}

Compared to the previous iteration of this analysis based on six years of data, more than 300,000 new events were detected and analyzed, see Table~\ref{tab:sample_properties}. In addition, the full sample of events was reprocessed within the scope of IceCube's \textit{Pass-2} campaign \citep{aartsenInSituCalibration2020} to apply latest detector calibrations also to archival data. In this campaign, the same event filtering, selections, and reconstructions are consistently applied to all data after May 2010 (IC79$-$IC86-2018), while IC59 is still treated separately in the analysis. More details can be found in \cite{stettner2021thesis}.

Five new events with reconstructed energies $E_{\mu , \mathrm{proxy}}\gtrsim \SI{200}{\tera\eV}$ are observed in the additional data-taking period. This energy approximately corresponds to a zenith-averaged signalness $S \gtrsim 0.5$: 

\begin{equation}
\label{eq:signalness}
S(E_{\mu}) = \frac{\Phi_\mathrm{signal}(E_{\mu})}{\Phi_\mathrm{signal}(E_{\mu}) + \Phi_\mathrm{background}(E_{\mu})} \gtrsim 0.5\,.
\end{equation} 

So, for these events the probability of the particle to belong to the differential astrophysical signal flux $\Phi_\mathrm{signal}$ exceeds $0.5$.

The event with the highest reconstructed energy from this additional period is the horizontal track event with ID II.32 (Table~\ref{tab:all_highe_events}) which has an energy of more than $E_{\mu , \mathrm{proxy}}\gtrsim  \SI{1.1}{\peta\eV}$, and has also been reported as realtime alert~\citep{gcn17569642}. The threshold $E_{\mu , \mathrm{proxy}}\gtrsim  \SI{200}{\tera\eV}$ to report individual events does not change significantly with the updated best-fit parameterization for the energy spectrum discussed in Section~\ref{sec:SPL}\footnote{With the updated spectral distribution as reported in this paper, a more precise threshold would correspond to an energy proxy of 185 TeV. For consistency with previously reported events, we keep the original threshold definition.}. The modified calibrations on the other hand, which include a $-4\%$-shift of the charge corresponding to a single photoelectron, do lead to changes of the reconstructed energies and directions of some of the events that have been reported in \cite{aartsenObservationCharacterizationCosmic2016}, shown in the rightmost three columns of Table \ref{tab:all_highe_events}. The impact on the reported quantities is small, especially for the directional reconstructions (most directions change by less than $0.1$ degree and stay well within their uncertainty ranges), but for single events with special topology, deviations in reconstructed energy can occur. Notably, such deviations arise for data taken in 2010 (IC79), where the information from the DeepCore strings is now excluded. As a consequence, six of the previously reported events fall below the threshold of $E_{\mu , \mathrm{proxy}}\gtrsim  \SI{200}{\tera\eV}$. Additionally, one event (ID6 in \cite{aartsenObservationCharacterizationCosmic2016}, IC79-season) does not pass the event selection anymore, because the IC79 season data as used in \cite{aartsenObservationCharacterizationCosmic2016} was based on an individual dedicated event selection, which has now been unified with the treatment of later seasons, where the event selection includes additional measures to reject cascade-like events more efficiently. These had not been applied to the IC79 data previously. For a detailed list of all events and the respective changes, see Table~\ref{tab:all_highe_events} in the supplementary material.

In order to calculate the expected number of events at the detector, a large sample of events is simulated, taking into account the following propagation steps: the propagation of neutrinos through matter, primary neutrino-nucleon interactions (CSMS cross-section:~\cite{coopersarkarHighEnergyNeutrino2011}), propagation of secondary particles through the ice (PROPOSAL code:~\cite{koehnePROPOSALToolPropagation2013}) and the production and propagation of Cherenkov photons in the detector medium \citep{Chirkin:2015kga}, CLSim code:~\cite{kopperCLSimCode2019}. They are modeled in detail before the final detector electronics and data acquisition are simulated. In addition, these simulations are repeated with altered assumptions on systematic uncertainties, e.g. scattering or absorption of the Cherenkov photons in the natural glacier and in the man-made boreholes. 

\section{Analysis method}
\label{sec:analysis_method}

The measurement of the energy spectrum of astrophysical muon neutrinos is performed via a forward-folding fit: the experimentally observed events are compared to the simulation-based expectation in a two-dimensional histogram as a function of reconstructed zenith angle and reconstructed muon energy. By maximizing the Poisson likelihood from Eq.~\ref{eq:poisson_LH}, the best-fitting flux hypothesis is obtained for the observed data $D$, which consists of a number of $n$ observed events per bin: 

\begin{align}
\label{eq:poisson_LH}
\mathcal{L}(D | \vec{\theta},\vec{\xi}) & = \prod_{\mathrm{bin}}^{N_{\mathrm{bins}}} p_{\mathrm{Poisson}}(n_{\mathrm{bin}}, \mu_{\mathrm{bin}}(\vec{\theta},\vec{\xi}))\,.
\end{align}

The expected number of events at the detector, $\mu$, is modeled as a function of signal parameters $\vec{\theta}$ and nuisance parameters $\vec{\xi}$. The latter absorb systematic uncertainties of the measurement, see Section~\ref{sec:nuisanceparameters}. 

The two-dimensional analysis histogram consists of $50$ energy bins equally spaced in $10^2 < \log_{10}(E_{\mu , \mathrm{proxy}} / \si{GeV}) < 10^7$ and $33$ bins spaced in reconstructed $\cos( \mathrm{Zenith})$ ranging from $85^\circ$ to $180^\circ$. The binning is identical to the one used in \cite{aartsenObservationCharacterizationCosmic2016} with $N_{\mathrm{bins}} = 1650$.

Four flux components are considered and the expected number of events is given as sum of these components per bin. First and second, atmospheric backgrounds are considered: \textit{conventional} atmospheric neutrinos, emerging from the decay of pions and kaons in cosmic ray induced air showers, and \textit{prompt} atmospheric neutrinos from the decay of heavy charmed hadrons in the same showers. For both contributions the prediction is updated with respect to \cite{aartsenObservationCharacterizationCosmic2016}, which used a prediction from \cite{Honda:Atmosflux2007}. The software \textit{Matrix Cascade Equation solver};\textit{ MCEq}~\citep{fedynitchCalculationConventionalPrompt2015} is employed assuming the flux model \textit{Gaisser-H4a}~\citep{gaisserSpectrumCosmicrayNucleons2012} for primary cosmic rays  and the hadronic interaction model \textit{Sibyll2.3c}~\citep{fedynitchHadronicInteractionModel2019}. Third, a sub-dominant contamination from atmospheric muons is expected and modeled based on a large sample of cosmic-ray air showers. The resulting model for this background flux is denoted as the \textit{muon-template}. It is simulated with the \textit{CORSIKA} package~\citep{corsika1998}, assuming the mentioned primary cosmic ray and hadronic interaction models. Fourth and most important, a signal component of astrophysical neutrinos is considered. It represents the cumulative flux from all sources of high-energy astrophysical neutrinos, which we assume to be isotropic. 

\begin{figure*}
    \centering
    \includegraphics[width=\textwidth]{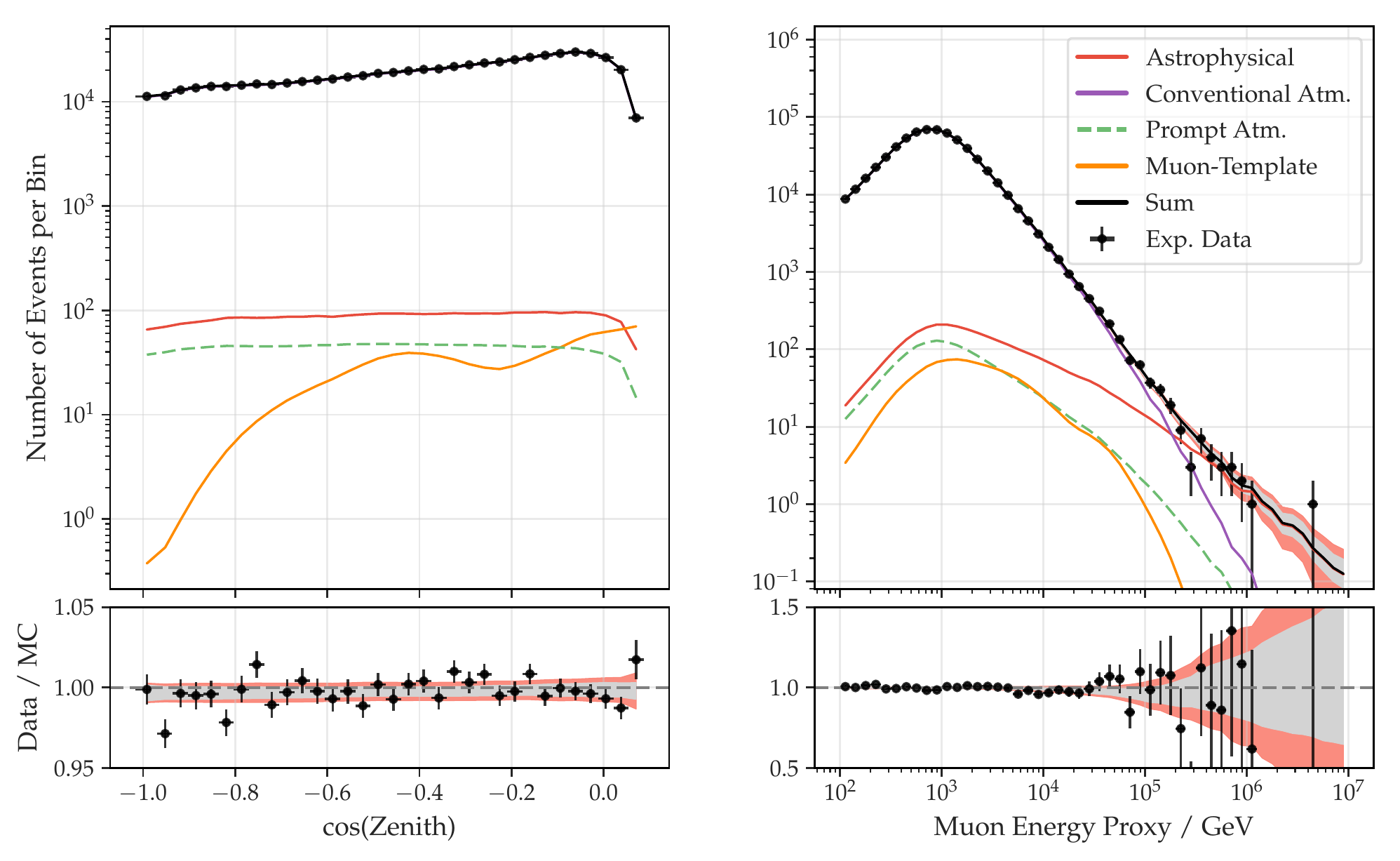}
    \caption{Single power-law model: Best-fit distributions, one-dimensional projection on reconstructed zenith angle and muon energy. The experimental data (black dots) are shown together with the best-fit expectation from simulation. Data taken in the IC59 detector configuration is kept in a separate analysis histogram. The conventional atmospheric component (purple) dominates the total flux for all zenith angles. Except for the highest energies, the line is thus hidden below the overall sum (black). The astrophysical component (red) is modeled as single power law. The prompt component is drawn at nominal prediction for visualization (green-dashed) although a zero best-fit normalization is obtained. The best-fit expectation for the remaining background of muons is shown by the orange line. The central $68\%$ range of the best-fit expectation is drawn as gray band. It is obtained by variation of all fit parameters according to their joint posterior distribution. The  red band additionally shows the statistical uncertainty of the simulated data.}
    \label{fig:bf_distributions_SPL}
\end{figure*}

\begin{figure}
    \centering
    \includegraphics[width=1.\columnwidth]{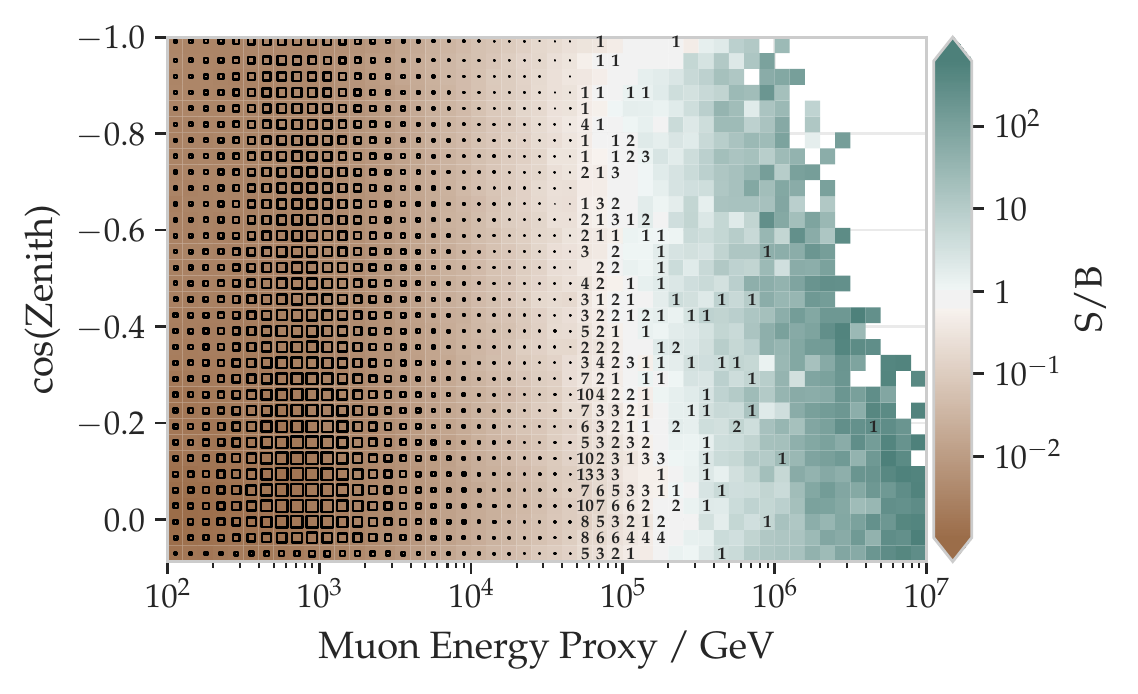}
    \caption{The color indicates the ratio of the signal (astrophysical) over the background (atmospheric and muonic) component expectation assuming the best-fit energy spectrum obtained in Section~\ref{sec:SPL}. The binning is equivalent to the two-dimensional histogram used in the analysis. For energies above $E_{\mu , \mathrm{proxy}}\gtrsim  \SI{50}{\tera\eV}$ the number of observed events per bin is indicated, in the bins below this threshold the number of events is proportional to the size of the black squares, with the maximum number of data events in a single bin being 3204.}
    \label{fig:expdata_signaloverbg}
\end{figure}

A wide range of parameterizations for the energy spectrum of the astrophysical component are investigated in Sections ~\ref{sec:SPL} and~\ref{sec:beyond_SPL}. The expected rate of events as a function of zenith and energy is visualized in Figure~\ref{fig:bf_distributions_SPL} for all four components assuming a single power law spectrum of astrophysical neutrinos. Figure~\ref{fig:expdata_signaloverbg} further illustrates the potential to measure astrophysical neutrinos at high energies by showing the expected ratio of astrophysical signal over background. 

\subsection{Updated Treatment of Systematic Uncertainties}
\label{sec:nuisanceparameters}
In order to absorb systematic uncertainties of the measurement, nuisance parameters are introduced in the fit. The first class of systematic uncertainties covers the detector response and reconstruction quality of IceCube. Optical properties of the natural ice and the ice in the man-made boreholes are considered as well as the optical detection efficiency. Compared to the last iteration of this analysis, the model of optical properties in the natural ice has been updated (\textit{Spice 3.2.1}:~\cite{rongenCalibrationIceCubeNeutrino2019}) and the impact of the ice in the boreholes (\textit{Unified hole-ice model}:~\cite{ellerUnifiedHoleiceModel2019}) is now taken into account as an additional source of uncertainty. For each of these systematic uncertainties, dedicated simulations have been performed and the impact on reconstructed energy and zenith has been parameterized. 

The second class of systematic uncertainties arises from the flux predictions of atmospheric neutrinos. Since the absolute normalization of the primary cosmic ray flux and the yield of neutrinos from cosmic-ray induced air showers are not known precisely, the absolute flux normalizations $\phi_{\mathrm{conv}}$ and $\phi_{\mathrm{prompt}}$ are free fit parameters. Additionally, the parameters $\Delta \gamma_{\mathrm{CR}}$ (CR spectral index shift) and $\lambda_{\mathrm{CR Model}}$, which is a prior-constrained linear interpolation between the  \textit{Gaisser-H4a} and \textit{GST-4gen} flux models~\citep{gaisserSpectrumCosmicrayNucleons2012,gaisserCosmicRayEnergy2013}, are introduced analogously to \cite{aartsenObservationCharacterizationCosmic2016} in order to cover uncertainties on the exact spectral shape of the primary cosmic ray flux. This shape affects all atmospheric flux components alike. Furthermore, we update the treatment of atmospheric flux uncertainties by adopting the scheme from \cite{barrUncertaintiesAtmosphericNeutrino2006}. By varying the relevant parameters $H$ (pions), $W$ (kaons), $Y$ (kaons) and $Z$ (kaons) within their estimated prior ranges and repeating the calculation of atmospheric fluxes, uncertainties from hadronic interaction models are covered by this suite of parameters. The parameters affecting the spectral shapes of the atmospheric fluxes are correlated to the absolute normalizations. For more details, see \ref{appendix:nuisance_parameters} and Figure \ref{fig:correlations_SPL} in the supplementary material. Finally, the absolute normalization of the sub-dominant muon contamination is included as a parameter and allowed to vary within its estimated prior range as well, which is a gaussian with a width of $0.5$.

\section{Results: Single power-law model}
\label{sec:SPL}

\begin{figure}
    \centering
    \includegraphics[width=1.\columnwidth]{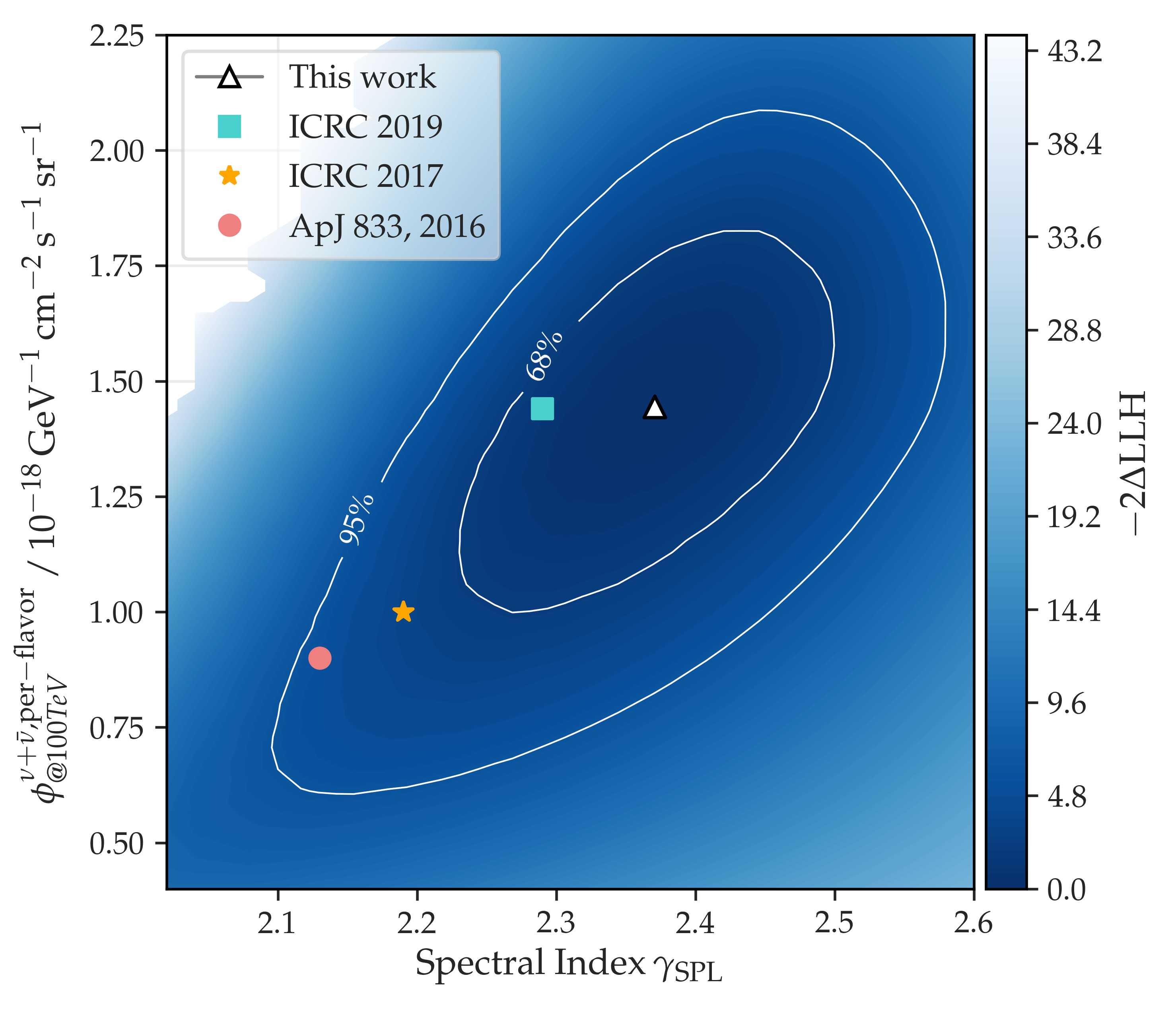}
    \caption{Single power-law model: Profile likelihood landscape as a function of spectral index and astrophysical normalization. The best-fit parameters are marked as white triangle. The turquoise square (\cite{stettner2019measurement},~$\gamma = 2.28^{+0.08}_{-0.09}$, $\Phi = 1.44^{+0.25}_{-0.24}$), the orange star (\cite{haackMeasurementDiffuseAstrophysical2018},~$\gamma = 2.19^{+0.10}_{-0.10}$, $\Phi = 1.01^{+0.26}_{-0.23}$) and the pink circle (\cite{aartsenObservationCharacterizationCosmic2016},~$\gamma = 2.13^{+0.13}_{-0.13}$, $\Phi = 0.90^{+0.30}_{-0.27}$) mark results of previous measurements.}
\label{fig:llh_contour_SPL}
\end{figure}

The result of the likelihood fit is shown as one-dimensional projection as solid black line in Figure~\ref{fig:bf_distributions_SPL} together with the experimental data. At energies above $E_{\mu , \mathrm{proxy}}\gtrsim  \SI{100}{\tera\eV}$, the excess of astrophysical neutrinos above the atmospheric components is clearly visible. Overall, the data are well described by the sum of atmospheric components and an astrophysical component following the standard paradigm of a single power-law energy spectrum. Figure~\ref{fig:pulls2d_bestfit} in the supplementary material shows the statistical pull for all bins in the two-dimensional histogram indicating no obvious mismatches. Taking the systematic and statistical uncertainty of the best-fit expectation into account, a $\chi^ 2 / \mathrm{(degrees~of~freedom)}$ for the single power-law fit is calculated to be $1.0$, resulting in a p-value of \SI{50}{\percent} and confirming that the fit result is a viable description of the measured data. The corresponding best-fit parameters of the astrophysical flux are listed in Table~\ref{tab:result_spl} and the profiled likelihood landscape of the two astrophysical signal parameters is shown in Figure~\ref{fig:llh_contour_SPL}. The sensitive energy range of the astrophysical measurement is determined by comparing the per-bin likelihood values of the best-fit hypothesis to the values obtained when repeating the fit assuming a background hypothesis. The true neutrino energy distribution is then weighted with these likelihood differences, and the central $90\%$-range of the obtained distribution is $E_{\nu} = 15\,\mathrm{TeV}$ to $5\,\mathrm{PeV}$. This energy range extends to lower energies than previous measurements, where this energy range extended from $E_{\nu} = 200\,\mathrm{TeV}$ to $8\,\mathrm{PeV}$ \citep{aartsenObservationCharacterizationCosmic2016}. This change is driven by the updated modeling of the conventional atmospheric flux in this energy region.

Compared to the previous analysis by  \cite{aartsenObservationCharacterizationCosmic2016}, a slightly softer spectral index of $\gamma_{\mathrm{SPL}} = 2.37_{-0.09}^{+0.09}$ is obtained. Figure~\ref{fig:llh_contour_SPL} shows the best fit points of the previous measurements, and the updates and changes between them are listed here as an overview. 
The measurements from \cite{aartsenObservationCharacterizationCosmic2016} and \cite{haackMeasurementDiffuseAstrophysical2018} are based on the same event selection and analysis method, with two years of additional data included in the latter. The changes between \cite{haackMeasurementDiffuseAstrophysical2018} and the results from \cite{stettner2019measurement} are mostly driven by the updated atmospheric background prediction. Additionally, the Barr parameters are introduced, resulting in more fit freedom at medium energies, and the sub-dominant effects of neutrino oscillations are considered. The updates to the detector simulations (Pass-2) also occurred in between these analyses, but have negligible effect on the fit result and uncertainty. Changes between the result reported in \cite{stettner2019measurement} and this analysis are a new event simulation including the effects of the hole-ice \citep{ellerUnifiedHoleiceModel2019} and the addition of the sub-dominant muon component.
A purely atmospheric hypothesis can be excluded  with very high significance at a $5.6 \sigma$ level. Note that this significance is smaller than previously reported results because of the updated more conservative treatment of systematic uncertainties as well as the observed softening of the spectral index.

\begin{table}[th]
	\centering
	{\small
		\begin{tabular}{lr}
			\hline
			{Astrophysical Norm. $\phi_{\mathrm{astro.}}/C_{\mathrm{units}} \, $} & $1.44_{-0.26}^{+0.25}$ \\
			{Spectral Index $\gamma_{\mathrm{SPL}}$} & $2.37_{-0.09}^{+0.09}$ \\
 			\hline
		\end{tabular}\\
	}
	\caption{Single power-law model: Best-fit parameters assuming a single power-law energy spectrum. The astrophysical normalization is given in units of $C_{\mathrm{units}} = 10^{-18}\,\mathrm{GeV}^{-1}\mathrm{cm}^{-2}\mathrm{s}^{-1}\mathrm{sr}^{-1}$. Confidence intervals ($68\%$) are constructed from one-dimensional profile likelihood scans employing Wilk's Theorem~\citep{wilksLargeSampleDistributionLikelihood1938}.}
\label{tab:result_spl}
\end{table}

\subsection{Prompt atmospheric neutrinos}
We find that the normalization of prompt atmospheric neutrinos is constrained less strongly compared to the last publication. This is related to the observed softening of the spectral index of the astrophysical neutrino flux, resulting in an overall more similar shape of the two flux components. For the calculation of a prompt limit, the dominating astrophysical flux in the regions below the sensitive energy range of this analysis poses a fundamental limitation to our ability to quantitatively constrain the sub-dominant flux of prompt neutrinos.

The best fit of the prompt normalization is still zero, independent of the different assumed astrophysical flux parameterizations discussed in Section~\ref{sec:beyond_SPL}. The prompt flux prediction from MCEq (Sibyll 2.3c; H4a) is used here compared to the ERS prediction~\citep{enbergPromptNeutrinoFluxes2008} in the last publication \citep{aartsenObservationCharacterizationCosmic2016}, which predict very similar fluxes. A more recent model \citep{bhattacharyaPerturbativeCharmProduction2015} predicts a prompt atmospheric flux component with about a factor of three smaller normalization.

It has been checked in detail that the measurement of the astrophysical component is not impacted by the prompt flux normalization: For example, we find a spectral index of $\gamma_{\mathrm{SPL}} = 2.33$, well within the quoted uncertainty range, if the likelihood fit is repeated with the prompt component fixed to its nominal prediction ($\phi_{\mathrm{prompt}}=1.0$). Also, the observed suppression of the prompt component occurs at similar strength when the likelihood fit is repeated while excluding energies above $\approx \SI{15}{\tera\eV}$, confining it to an energy region dominated by conventional atmospheric neutrinos and well below the sensitive energy range for astrophysical neutrinos. The exact reason for this non-observation remains an open question and an updated limit on the prompt flux normalization with respect to \cite{aartsenObservationCharacterizationCosmic2016} is not computed here. 

\section{Results: Beyond the single power law}
\label{sec:beyond_SPL}

Power-law energy spectra are well motivated from the assumed acceleration mechanisms of cosmic rays, but they extrapolate over large energy ranges and potential structures in the energy spectrum can thus not be identified. In this section, a number of parameterizations beyond a single power law are compared to the experimental data. Parameterizations with more than three signal parameters are however not considered, because the statistics of observed events with high signalness is too low to constrain more fit parameters. 

\subsection{Power law with cutoff}
The first natural extension to the single power law would be a cutoff in the energy spectrum, e.g.~introduced if an astrophysical source of cosmic rays (and neutrinos) reaches its maximum energy. The flux parameterization given in Eq.~\ref{eq:models_SPLwcutoff} extends the single power law with an exponentially decaying term. The cut-off neutrino energy $E_{\text{cutoff}}$ is consequently added as a third signal parameter in the likelihood fit:

\begin{flalign}
\label{eq:models_SPLwcutoff}
& \Phi_{\mathrm{astro.}}^{\nu_\mu+\bar{\nu}_\mu}(E_\nu) = & \hfill \\ 
& \phi_{\text{cutoff}}  \times  \left(\frac{E_\nu}{\SI{100}{\tera\eV}}\right)^{-\gamma_{\text{cutoff}}}  \times e^{\left(\frac{-E_\nu}{E_{\text{cutoff}}}\right)}\nonumber\,.
\end{flalign}

Table~\ref{tab:result_cutoff} lists the obtained fit result using this parameterization: While the astrophysical normalization does not change strongly, a hard spectral index of $\gamma_{\text{cutoff}} = 2.0_{-0.28}^{+0.22}$ and a cut-off energy of $E_{\text{cutoff}} = 1.25_{-0.56}^{+1.72} \, \si{\peta\eV}$ is found. Compared to the single power-law hypothesis, the fit improves by $2\Delta\mathrm{LLH} = 4.24$. The probability to randomly achieve any such improvement by introducing $E_{\text{cutoff}}$ corresponds to a p-value of $\mathrm{p}(>2\Delta\mathrm{LLH} | \mathrm{SPL})=6.1 \%$, which is calculated from pseudo-experiments obtained from Monte-Carlo simulations.

\begin{table}[th]
	\centering
	{\small
		\begin{tabular}{lr}
			\hline
			{Astrophysical Norm. $\phi_{\text{cutoff}} / \, C_{\mathrm{units}}$} & $1.64_{-0.36}^{+0.39}$ \\
			{Spectral Index $\gamma_{\text{cutoff}}$} & $2.0_{-0.28}^{+0.22}$ \\
			{Cut-off Energy $E_{\text{cutoff}} / \, \si{\peta\eV}$} & $1.25_{-0.56}^{+1.72} $\\
			\hline
			{Significance over SPL} & $2\Delta\mathrm{LLH} = 4.24 $\\
			\multicolumn{2}{r}{$\mathrm{p}(>2\Delta\mathrm{LLH} | \mathrm{SPL})=6.1 \%$} \\
			\hline
			
		\end{tabular}\\
	}
	\caption{Single power law with cutoff: Best-fit parameters. Confidence intervals ($68\%$) are constructed from one-dimensional profile likelihood scans employing Wilk's Theorem.}
	\label{tab:result_cutoff}
\end{table}

\subsection{Log-Parabola Model}
\label{sec:logparabola}
Similarly to the cut-off hypothesis, the log-parabola model, widely used in gamma-ray astronomy, extends the single power law and allows for curvature of the spectrum. Its parameterization is given in Eq.~\ref{eq:models_LogParabola} and Table~\ref{tab:result_logparabola} lists the obtained best-fit parameters of the likelihood fit with an astrophysical component following this model. Again, a hard spectral index of $\alpha_{\mathrm{LogParab.}} = 2.03_{-0.31}^{+0.22}$ is obtained, and the best fit of the curvature parameter is $\beta_{\mathrm{LogParab.}} = 0.45_{-0.22}^{+0.29}$. Compared to the single power-law hypothesis, which corresponds to $\beta_{\mathrm{LogParab.}} = 0$, the description of the experimental data is improved by $2\Delta\mathrm{LLH} = 6.82$. Analogously to the treatment for the cut-off hypothesis, this can be translated to a p-value of $\mathrm{p}(>2\Delta\mathrm{LLH} | \mathrm{SPL})=1.3 \%$.

\begin{flalign}
\label{eq:models_LogParabola}
&\Phi_{\mathrm{astro.}}^{\nu_\mu+\bar{\nu}_\mu}(E_\nu) =  & \hfill \\
& \phi_{\mathrm{LogParab.}}  \times \left(\frac{E_\nu}{\SI{100}{\tera\eV}}\right)^{-\alpha_{\mathrm{LogParab.}} -\beta_{\mathrm{LogParab.}} \log(\frac{E_\nu}{\SI{100}{\tera\eV}})}\nonumber
\end{flalign}

\begin{table}[htb]
	\centering
	{\small
		\begin{tabular}{lr}
			\hline
			{Log-parabola Norm. $\phi_{\mathrm{LogParab.}} / \, C_{\mathrm{units}} $} & $1.79_{-0.38}^{+0.40}$ \\
			{Spectral Index $\alpha_{\mathrm{LogParab.}}$} & $2.03_{-0.31}^{+0.22}$ \\
			{Curvature parameter $\beta_{\mathrm{LogParab.}}$} & $0.45_{-0.22}^{+0.29}$ \\
			\hline
			{Significance over SPL} & $2\Delta\mathrm{LLH} = 6.82 $\\
			\multicolumn{2}{r}{$\mathrm{p}(>2\Delta\mathrm{LLH} | \mathrm{SPL})=1.3 \%$} \\
			\hline
		\end{tabular}\\
	}
	\caption{Log-Parabola model: Best-fit parameters. Confidence intervals ($68\%$) are constructed from one-dimensional profile likelihood scans employing Wilks' Theorem.}
	\label{tab:result_logparabola}
\end{table}

\subsection{Piece-wise Parameterization}
\label{sec:piecewise}
In order to overcome the limitations of the parameterizations discussed in previous sections, a 'piece-wise' model is introduced. Here, the energy spectrum is described as sum of power laws with a fixed spectral index ($\gamma=2.0$) in pre-defined, fixed segments of neutrino energy. This allows for measuring the flux strength in a well-defined range of neutrino energy and enables easy comparison to predictions from the literature and to other measurements. The total flux strength is then given by Eq. \ref{eq:models_piecewise}, where the flux normalizations per bin $\phi_{\mathrm{piece}}^{i}$ are fit parameters and $E_{\mathrm{low}}^{i}$ and $E_{\mathrm{high}}^{i} $ form the bounds of bin $i$.

\begin{flalign}
\label{eq:models_piecewise}
\Phi_{\mathrm{astro.}}^{\nu_\mu+\bar{\nu}_\mu}(E_\nu) =  &  \sum_{i}^{\mathrm{pieces}} \chi(E_\nu) \cdot \phi_{\mathrm{piece}}^{i} \cdot  \left(\frac{E_\nu}{\SI{100}{\tera\eV}}\right)^{-2.0} & \hfill \\
\chi(E_\nu) = &    \begin{cases} 
  										1   & \text{if} \, E_{\mathrm{low}}^{i} <  E_\nu < E_{\mathrm{high}}^{i} \\
  										0  & \text{else}
  					 \end{cases} & \nonumber
\end{flalign}

\begin{figure}
    \centering
    \includegraphics[width=1.\columnwidth]{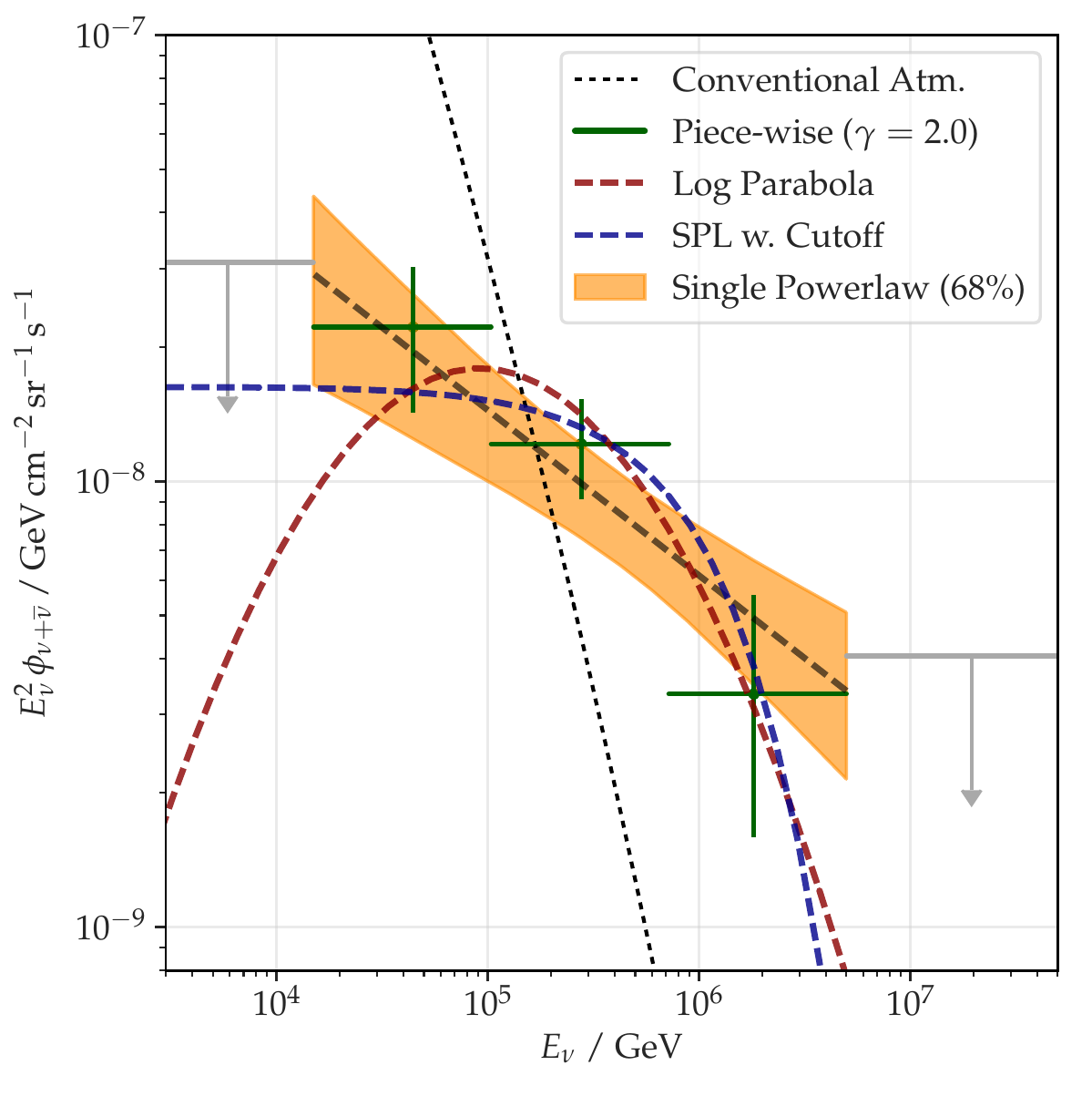}
    \caption{Summary of best-fit models for the astrophysical neutrino flux. The bins from the piece-wise unfolding are marked in green and in gray wherever only upper limits are calculated. The single power law band is drawn in the sensitive energy range as defined in Section \ref{sec:SPL}.  All models with more degrees of freedom than the single power law show a trend from a hard spectral shape at medium energies to a softer spectrum at highest energies.}
    \label{fig:results_allparameterizations}
\end{figure}

Prior to performing the fit on the experimental data, the energy ranges of the segments were defined to be equally spaced in log-energy spanning the sensitive energy range of the astrophysical measurement (see Section~\ref{sec:SPL}) with three segments. Additionally, one segment above and below have been added respectively to cover the full energy range. The full parameterization of the astrophysical flux is given in Eq.~\ref{eq:models_piecewise}, and the energy ranges and obtained best-fit normalizations $\phi_{\mathrm{piece}}^{i}$ are listed in Table~\ref{tab:result_pieces}. Figure~\ref{fig:results_allparameterizations} visualizes the obtained flux measurement of the piece-wise parameterization together with the results of the single power law, power law with cut-off and log-parabola models. In all models beyond the single power law, hints for a softening of the spectral shape as a function of energy are found. 

\begin{table}[h!]
		\centering
		{\small
			\begin{tabular}{llr}
				\hline
				{} & Energy Range ($E_\nu$) & Norm. $\phi_{\mathrm{piece}}^{i} / C_{\mathrm{units}}$ \\
				\hline 
				Piece 1 & $\SI{100}{\giga\eV} - \SI{15}{\tera\eV}$ & $^{\dagger}0.0^{+3.1} $ \\
				Piece 2 & $ \SI{15}{\tera\eV} - \SI{104}{\tera\eV}$ &  $ 2.22^{+0.8}_{-0.8} $ \\
				Piece 3 & $ \SI{104}{\tera\eV} - \SI{721}{\tera\eV}$ & $1.21^{+0.32}_{-0.31}$ \\
				Piece 4 & $ \SI{721}{\tera\eV} - \SI{5}{\peta\eV}$ & $0.33^{+0.22}_{-0.18}$ \\
				Piece 5 & $ \SI{5}{\peta\eV} - \SI{100}{\peta\eV}$ & $^{\dagger}0.0^{+0.41}$\\
				\hline
			\end{tabular}\\
		}
		\caption{Piece-wise parameterization: Energy ranges and result of the likelihood fit. 
		Note that all piece-wise normalizations are optimized simultaneously in the fit, i.e. correlations between the segments are fully taken into account. The given $68.27\%$ uncertainty ranges are obtained from one-dimensional profile likelihood scans.\\
		$^{\dagger}$Piece 1 and 5 have been added to cover the full energy range, here, upper limits ($90\%$ CL) are computed.}
		\label{tab:result_pieces}
\end{table}

\subsection{Flux predictions for specific source classes}
\label{sec:astromodels}
Besides the wide range of generic parameterizations for the energy spectrum discussed in the sections above, it is also possible to compare the experimental data to source-class specific flux predictions directly. The total astrophysical flux may originate from multiple source-classes, it is thus not expected that a single flux prediction can fully explain the observed data. Instead, we model the total astrophysical component as sum of the predicted energy spectrum model times a free normalization $\phi_{\mathrm{model}}$ and a single power law to cover other potential flux contributions:
\begin{flalign}
\label{eq:models_theorypredictions}
\Phi_{\mathrm{astro.}}^{\nu_\mu+\bar{\nu}_\mu}(E_\nu) = & \phi_{\mathrm{model}} \times \mathrm{Model}(E_{\nu}) \\ 
 & + \phi_{\mathrm{SPL}} \times \left(\frac{E_\nu}{\SI{100}{\tera\eV}}\right)^{-\gamma_{\mathrm{SPL}}}.\nonumber
\end{flalign}

A representative set of different source-class specific predictions have been selected, focusing on predictions not already covered by the performed test of a single power law, and  including variations of the benchmark models shown in the publications (see Table~\ref{tab:modelfits_free}). All these predictions model the cumulative expected flux at Earth for the given source class. The obtained fit results using these predictions are listed in Table~\ref{tab:modelfits_free}. The test-statistic $TS_{\mathrm{free\,model}}$ from Eq.~\ref{eq:TS_theorytests_free} compares the best-fit result including the additional component of the source-class specific flux prediction to the hypothesis of only a single power-law. That is, $TS_{\mathrm{free\,model}} = 0$ implies that the description of the experimental data can not be improved with an additional contribution from the model prediction and the fit instead prefers the  single power-law model. For these cases, upper limits on the model normalization are computed at $90\%$ CL employing Wilk's Theorem. 

\begin{align}
\label{eq:TS_theorytests_free}
& TS_{\mathrm{free\,model}} = & \\ \nonumber
& -2\times \log \left( \frac{\mathcal{L}(D | \hat{\phi}_{\mathrm{model}} , \hat{\phi}_{\mathrm{SPL}}, \hat{\gamma}_{\mathrm{SPL}}, \hat{\vec{\xi}})}{\mathcal{L}(D | \phi_{\mathrm{model}}\equiv 0.0 , \phi_{\mathrm{SPL}}, \gamma_{\mathrm{SPL}}, \vec{\xi})} \right) &
\end{align}

Similar to the results using the generic parameterizations for the energy spectrum, we find that model predictions with a softening of the spectral shape in the energy range $\SI{100}{\tera\eV} - \SI{1}{\peta\eV}$ describe the experimental data better than the single power law. For example, the addition of spectral components as predicted by the models from \cite{sennoChokedJetsLowLuminosity2016} and \cite{liuCanWindsDriven2018} ($\Gamma = 2.3$) are favored compared to the pure single power-law hypothesis, with test statistics reaching $-TS_{\mathrm{free\,model}} > 4$. For these two models, the best-fit model normalizations are multiples of the model prediction, substantially reducing the strength of the single power-law component, while neither model is claiming to account for the entire diffuse emission. When fixing the model normalization ($\Phi_{\mathrm{model}} = 1.0$ in equation \ref{eq:models_theorypredictions}) and comparing to the single power-law model, only the models from \cite{liuCanWindsDriven2018} and \cite{sennoChokedJetsLowLuminosity2016} result in negative test statistics, indicating a small but insignificant preference of those model predictions \cite{stettner2021thesis}. For every other model in Table \ref{tab:modelfits_free} which is yielding negative test statistics when allowing a free model normalization, the fitted normalization is smaller than the nominal prediction in the original publication.

Models in Table \ref{tab:modelfits_free} predicting some spectral hardening in the considered energy range, like \cite{biehlCosmicRayNeutrinoEmission2018}, \cite{muraseDiffuseNeutrinoIntensity2014} and \cite{padovaniSimplifiedViewBlazars2015} are mildly disfavored. While this does not disfavor them as source models for individual neutrino sources, they are less likely to be main contributors to the overall diffuse astrophysical neutrino flux.

\begin{table*}[ht]
\centering
\begin{tabular}{llcccccc}
\hline
Model & Variation & $TS_{\mathrm{free\,model}} $ & $\phi_{\mathrm{model}}$ & $\phi_{\mathrm{astro.}}^{\mathrm{SPL}} $ & $\gamma_{\mathrm{SPL}}$ & UL $ \phi_{\mathrm{model}}^{90\%}$ \\
\hline
\cite{biehlCosmicRayNeutrinoEmission2018} (GRB) 
& Sum model A & $ 0.00 $&  $ 0.00 $ & $1.44 $ & $2.37 $ & $ {0.19} $  \\ 
& Sum model B & $ 0.00 $&  $ 0.00 $ & $1.44 $ & $2.37 $ & $ {4.92} $  \\ 
\cite{sennoExtragalacticStarformingGalaxies2015} (SFG w. HNe) 
& Diffusion $\propto E^{\frac{1}{2}}$ & $ -0.14 $&  $ 0.12_{-0.34}^{+0.27}$ & $1.12 $ & $2.40 $ &  -  \\ 
& Diffusion $\propto E^{\frac{1}{3}}$ & $ -0.41 $&  $ 0.23_{-0.37}^{+0.29}$ & $0.87 $ & $2.42 $ &  -  \\ 
\cite{muraseDiffuseNeutrinoIntensity2014} (AGN inner Jets) 
& $\Gamma=2.0$, Blazar & $ 0.00 $&  $ 0.00 $ & $1.44 $ & $2.37 $ & $ {0.48} $  \\ 
& $\Gamma=2.0$, Torus  & $ 0.00 $&  $ 0.00 $ & $1.44 $ & $2.37 $ & $ {0.58} $  \\ 
& $\Gamma=2.3$, Blazar & $ 0.00 $&  $ 0.00 $ & $1.44 $ & $2.37 $ & $ {0.48} $  \\ 
& $\Gamma=2.3$, Torus  & $ 0.00 $&  $ 0.00 $ & $1.44 $ & $2.37 $ & $ {0.27} $  \\ 
\cite{liuCanWindsDriven2018} (AGN winds) 
& CR ($\Gamma=2.1$) & $ -0.98 $&  $ 0.87_{-0.88}^{+0.49}$  & $0.47 $ & $2.47 $ &  -  \\ 
& CR ($\Gamma=2.3$) & $ -4.12 $&  $ 14.3_{-5.97}^{+3.61}$ & $0.12 $ & $2.04 $ &  -  \\ 
\cite{padovaniSimplifiedViewBlazars2015} (BL-Lac) 
& $\frac{F_{\nu}}{F_{\gamma}}=0.3$ & $ 0.00 $&  $ 0.00 $ & $1.44 $ & $2.37 $ & $ {0.27} $  \\ 
& $\frac{F_{\nu}}{F_{\gamma}}=0.8$ & $ 0.00 $&  $ 0.00 $ & $1.44 $ & $2.37 $ & $ {0.1} $  \\ 
\cite{kimuraNeutrinoCosmicrayEmission2015} (lowL-AGN) 
& Model B1 & $ -1.69 $ &  $ 0.33_{-0.25}^{+0.23}$ & $0.89 $ & $2.46 $ &  -          \\ 
& Model B4 & $ 0.00  $ &  $ 0.00 $                & $1.44 $ & $2.37 $ & $ {0.24} $  \\ 
\cite{BiehlTDStarsAsCRandNuOrigin2018} (TDE) 
& No variations & $ 0.00 $&  $ 0.00 $ & $1.44 $ & $2.37 $ & $ {0.63} $  \\ 
\cite{tavecchioHighenergyCosmicNeutrinos2014} (lowL-BLLac) 
& No variations & $ -1.74 $&  $ 0.32_{-0.24}^{+0.22}$ & $0.82 $ & $2.47 $ &  -  \\ 
\cite{sennoChokedJetsLowLuminosity2016} (GRB w. choked Jets) 
& No variations & $ -4.36 $&  $ 2.6_{-1.16}^{+0.4}$ & $0.00 $ & - &  -  \\ 
\hline
\end{tabular}\\
\caption{Results of the likelihood fits with an additional astrophysical component following a source-class specific flux prediction. The normalization is added as an additional fit parameter. A negative log-likelihood difference indicates that the data is better described with the additional component compared to the single power-law model. The last column shows the $90\%$ CL (Wilk's Theorem) upper limit if the model normalization is fitted to zero. Tested source classes include different GRB and AGN scenarios, star forming galaxies (SFG) with hypernovae  (HNe), and TDEs. Low luminosity models are preceded by \enquote{lowL}.\label{tab:modelfits_free}}
\end{table*}

\section{Discussion and Outlook}
\label{sec:discussion}

\begin{figure*}[htbp]
    \centering
    \includegraphics[width=1.\textwidth]{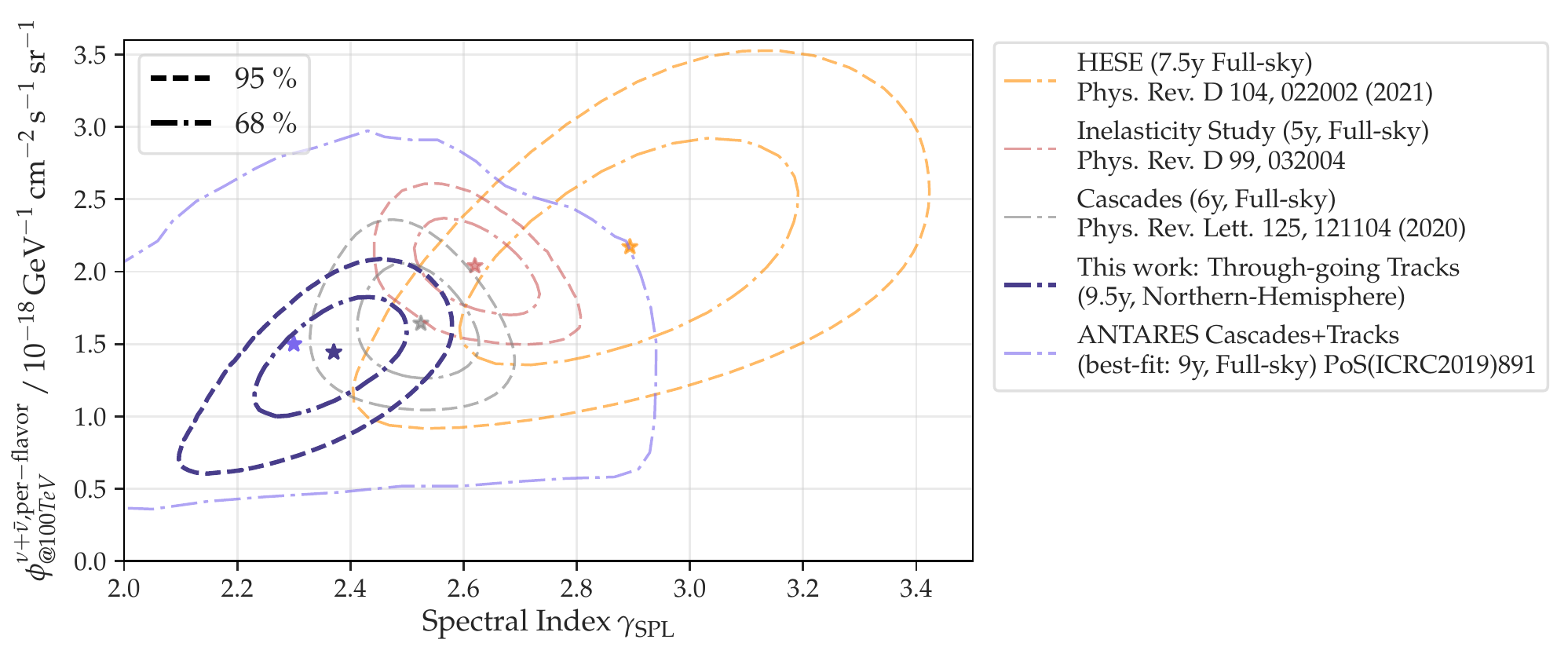}
    \caption{Summary of astrophysical neutrino-flux measurements. Best-fit parameters and uncertainty contours for the single power-law hypothesis are drawn for studies based on high-energy starting events~\citep{Abbasi:2020jmhHESE75}, cascade-like events~\citep{aartsenCharacteristicsDiffuseAstrophysical2020}, and an inelasticity study~\citep{Aartsen:2018vezInelasticity} by IceCube. ANTARES observes a mild excess of events over the expected atmospheric backgrounds in a combined study of tracks and cascades \citep{fuscoStudyHighenergyNeutrino2019}.}
    \label{fig:comparison_SPL}
\end{figure*}

We have presented an updated measurement of the astrophysical muon-neutrino flux from the northern celestial sky. The last measurement from \cite{aartsenObservationCharacterizationCosmic2016} observed a flux compatible with a single power law described by a spectral index of $\gamma = 2.13^{+0.13}_{-0.13}$. Our update consists of more than three years of additional data (roughly doubling statistics), a re-processing of the full data with latest calibration and filtering standards (Pass-2) and an updated treatment of systematic uncertainties (detector effects and atmospheric fluxes). Assuming a single power-law energy spectrum, we find a spectral index of $2.37_{-0.09}^{+0.09}$. This is in agreement with but slightly softer than earlier iterations of this analysis. This change is partly caused by the updated atmospheric flux models and uncertainty treatment but also by updated detector simulations of photon detection as well as the added data. In addition, we tested parameterizations beyond the single power law for the first time and find hints for a softening of the spectral shape as a function of energy at the two-sigma confidence level (Cut-off, Log-parabola and piece-wise model). Figure~\ref{fig:comparison_SPL} shows the result of the spectral fit in comparison to other measurements of the diffuse astrophysical flux by IceCube \citep{Aartsen:2018vezInelasticity, aartsenCharacteristicsDiffuseAstrophysical2020,  Abbasi:2020jmhHESE75} and the mild excess over expected atmospheric backgrounds observed by ANTARES \citep{fuscoStudyHighenergyNeutrino2019}. In the energy range where all referred analyses are sensitive, the observed event rates agree well with each other. The analyses themselves are based on event samples of varying statistical sizes, covering different energies and neutrino flavors. The advantages of the different analysis methods and the relations between the spectral results shown in Figure~\ref{fig:comparison_SPL} are discussed in detail in \cite{Abbasi:2020jmhHESE75}. With continued data taking of IceCube, it is expected that these measurements can be further improved in the future. Furthermore, with the future IceCube-Gen2 Observatory \citep{AartsenGen2:2020}, we expect a substantial increase of exposure by a factor of $\approx6$, which will improve the statistics in the here probed energy range and also allow for extensions of the energy range to higher energies. A further goal of the collaboration is combining the measured astrophysical fluxes from different detection channels into a single consistent analysis based on a global fit of the data. 


\section*{Acknowledgements}
The IceCube collaboration acknowledges the significant contributions to this manuscript from Philipp F\"urst, J\"oran Stettner, and Christopher Wiebusch. We  acknowledge  the  support  from the  following  agencies: USA {\textendash} U.S. National Science Foundation-Office of Polar Programs,
U.S. National Science Foundation-Physics Division,
U.S. National Science Foundation-EPSCoR,
Wisconsin Alumni Research Foundation,
Center for High Throughput Computing (CHTC) at the University of Wisconsin{\textendash}Madison,
Open Science Grid (OSG),
Extreme Science and Engineering Discovery Environment (XSEDE),
Frontera computing project at the Texas Advanced Computing Center,
U.S. Department of Energy-National Energy Research Scientific Computing Center,
Particle astrophysics research computing center at the University of Maryland,
Institute for Cyber-Enabled Research at Michigan State University,
and Astroparticle physics computational facility at Marquette University;
Belgium {\textendash} Funds for Scientific Research (FRS-FNRS and FWO),
FWO Odysseus and Big Science programmes,
and Belgian Federal Science Policy Office (Belspo);
Germany {\textendash} Bundesministerium f{\"u}r Bildung und Forschung (BMBF),
Deutsche Forschungsgemeinschaft (DFG),
Helmholtz Alliance for Astroparticle Physics (HAP),
Initiative and Networking Fund of the Helmholtz Association,
Deutsches Elektronen Synchrotron (DESY),
and High Performance Computing cluster of the RWTH Aachen;
Sweden {\textendash} Swedish Research Council,
Swedish Polar Research Secretariat,
Swedish National Infrastructure for Computing (SNIC),
and Knut and Alice Wallenberg Foundation;
Australia {\textendash} Australian Research Council;
Canada {\textendash} Natural Sciences and Engineering Research Council of Canada,
Calcul Qu{\'e}bec, Compute Ontario, Canada Foundation for Innovation, WestGrid, and Compute Canada;
Denmark {\textendash} Villum Fonden and Carlsberg Foundation;
New Zealand {\textendash} Marsden Fund;
Japan {\textendash} Japan Society for Promotion of Science (JSPS)
and Institute for Global Prominent Research (IGPR) of Chiba University;
Korea {\textendash} National Research Foundation of Korea (NRF);
Switzerland {\textendash} Swiss National Science Foundation (SNSF);
United Kingdom {\textendash} Department of Physics, University of Oxford.

\newpage
\begin{appendices}
\section{Supplementary Material}
\label{sec:supp_higheevents}

\subsection{Barr-Treatment of Atmospheric Uncertainties}
\label{appendix:nuisance_parameters}

\begin{figure}[hb]
    \centering
    \includegraphics[width=1.\columnwidth]{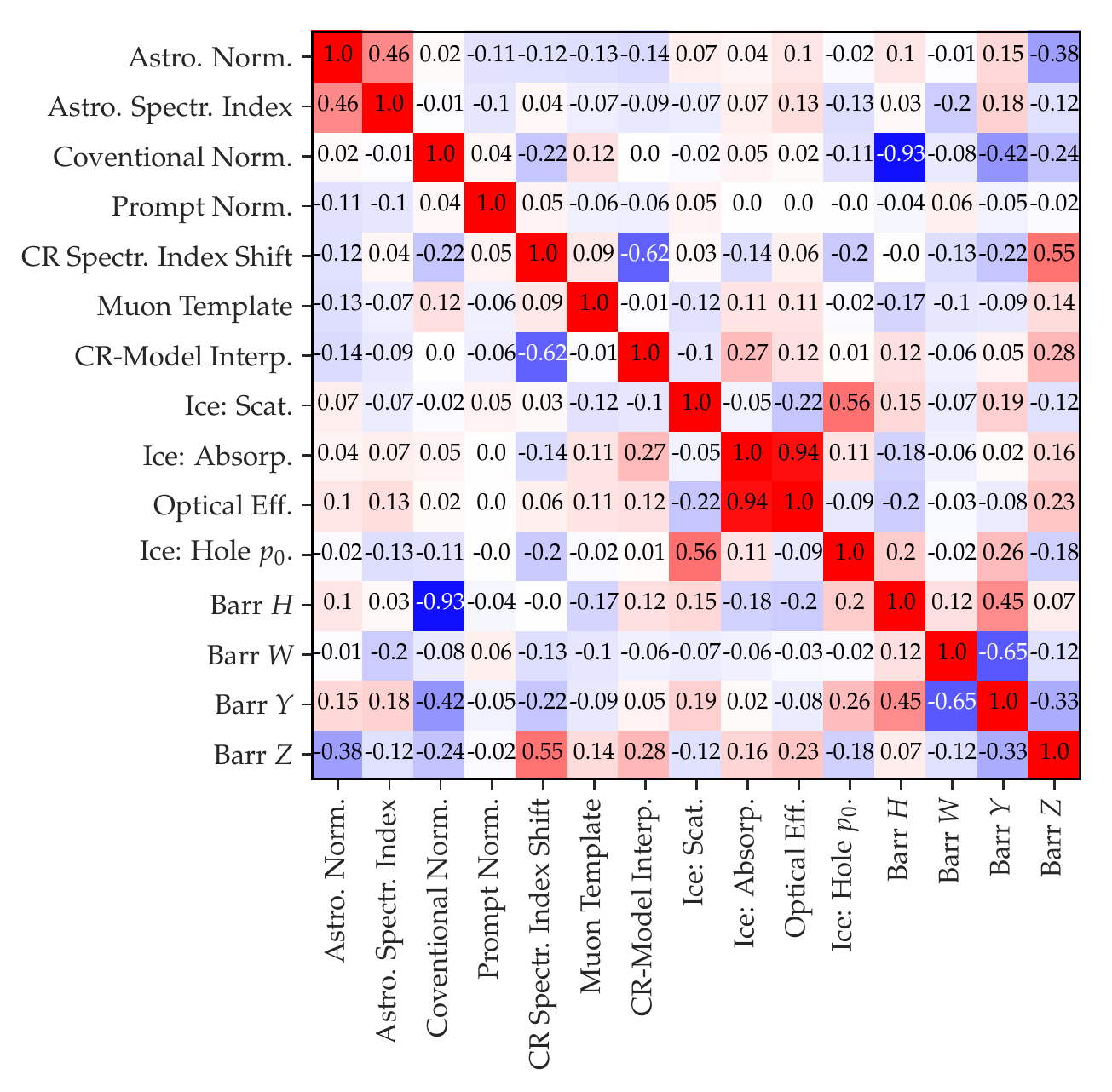}
    \caption{Pearson correlation coefficients between the signal and nuisance parameters are shown for the parameters of the single power-law fit.}
    \label{fig:correlations_SPL}
\end{figure}

The nuisance parameters are included in the fit to cover the systematic uncertainties affecting this measurement, with the goal of measuring an unbiased result. The scheme from \cite{barrUncertaintiesAtmosphericNeutrino2006} was adopted to cover atmospheric flux uncertainties. Previous analyses used a parameter describing the ratio between the integrated neutrino fluxes arising from kaon and pion decays, respectively \citep{aartsenObservationCharacterizationCosmic2016, haackMeasurementDiffuseAstrophysical2018}. In principle, the Barr scheme allows for an uncertainty in the production yield of each individual meson, for example pions and antipions. Different from a global scaling of these production yields, each parameter describes uncertainties in a specific region of meson production phase space, with the goal of having different parameters for regions dominated by different physical effects and with different experimental coverage. Since the total $\nu + \nubar$ flux is measured in this analysis, the parameters for mesons and antimesons can be combined into single parameters here. Flux gradients are then calculated from flux predictions obtained with different parameter values, and the Barr-parameters in the fit then scale this gradient to obtain a flux prediction depending on parameter value. Since they affect the neutrino production, the Barr-parameters are correlated to the absolute normalization of the conventional atmospheric flux, but crucially also introduce energy-dependent flux variations \citep{stettner2021thesis}. The correlations between the nuisance parameters are shown in Figure \ref{fig:correlations_SPL}.

\begin{figure}[hb]
    \centering
    \includegraphics[width=1.\columnwidth]{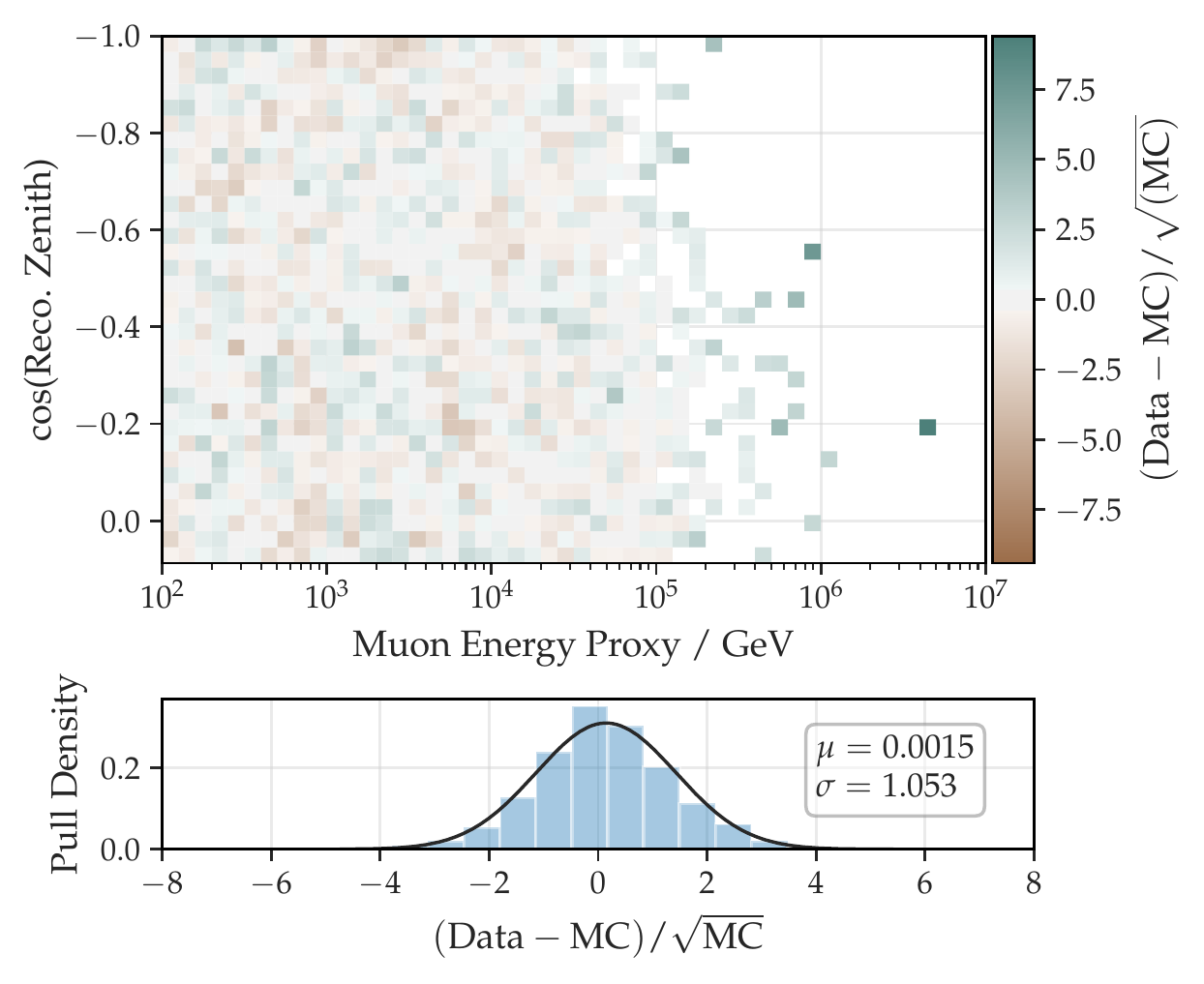}
    \caption{The upper figure shows the statistical pull per bin between the experimental data and the MC expectation assuming the best-fit energy spectrum obtained in Section~\ref{sec:SPL}. The lower figure shows the pull density distribution for the 1048 analysis bins containing data events.}
    \label{fig:pulls2d_bestfit}
\end{figure}

\begin{table*}
\centering
\begin{longtable}{llllllccc}
			\hline
ID & MJD & Energy / TeV & Signalness & Decl. / deg & R.A. / deg &  $\Delta_{\mathrm{Energy}}$ / TeV &  $\Delta_{Decl.} $/ deg&  $\Delta_{RA}$ / deg \\
\hline
I.4$^{ \dagger }$ & $55370.7 $ & $ 150 $ & --- &               --- & --- & $-110$ & --- & --- \\
I.6$^{ \ddagger }$ & $55421.5 $ & --- & --- & --- & --- & --- & --- & --- \\
I.13$^{ \dagger }$ & $55722.4 $ & $ 180 $ & --- &               --- & --- & $-30$ & --- & --- \\
I.15$^{ \dagger }$ & $55896.9 $ & $ 110 $ & --- &               --- & --- & $-190$ & --- & --- \\
I.17$^{ \dagger }$ & $56063.0 $ & $ 180 $ & --- &               --- & --- & $-20$ & --- & --- \\
I.19$^{ \dagger }$ & $56211.8 $ & $ 200 $ & --- &               --- & --- & $-10$ & --- & --- \\
I.28$^{ \dagger }$ & $57049.5 $ & $ 190 $ & --- &               --- & --- & $-20$ & --- & --- \\
\hline
II.1$^{\mathrm{IC59}}$ & $55056.7$ & $480$ & $0.78$ &    $1.23_{-0.22}^{+0.18}$ &     $29.51_{-0.38}^{+0.40}$ &     --- & --- & --- \\
II.2$^{\mathrm{IC59}}$ & $55141.1$ & $250$ & $0.52$ &    $11.74_{-0.38}^{+0.32}$ &     $298.21_{-0.57}^{+0.53}$ &     --- & --- & --- \\
II.3 & $55355.5$ & $350$ & $0.72$ &    $22.63_{-2.15}^{+2.72}$ &     $346.29_{-2.33}^{+2.11}$ &     $10$ & $-0.95$ & $1.36$ \\
II.4 & $55387.5$ & $200$ & $0.53$ &    $21.22_{-1.73}^{+2.53}$ &     $307.31_{-2.50}^{+2.77}$ &     $-30$ & $0.22$ & $0.35$ \\
II.5 & $55464.9$ & $390$ & $0.74$ &    $13.48_{-0.24}^{+0.32}$ &     $266.26_{-0.39}^{+0.55}$ &     $-70$ & $0.08$ & $-0.03$ \\
II.6 & $55478.4$ & $600$ & $0.84$ &    $11.03_{-0.36}^{+0.30}$ &     $331.13_{-0.48}^{+0.39}$ &     $-60$ & $-0.06$ & $0.05$ \\
II.7 & $55497.3$ & $850$ & $0.88$ &    $0.34_{-0.14}^{+0.16}$ &     $88.86_{-0.20}^{+0.29}$ &     $-100$ & $-0.16$ & $-0.09$ \\
II.8 & $55513.6$ & $460$ & $0.79$ &    $3.17_{-0.27}^{+0.31}$ &     $285.56_{-0.32}^{+0.48}$ &     $-60$ & $0.02$ & $-0.39$ \\
II.9 & $55589.6$ & $210$ & $0.55$ &    $1.16_{-0.29}^{+0.20}$ &     $307.79_{-0.51}^{+0.40}$ &     $-30$ & $0.13$ & $0.08$ \\
II.10 & $55702.8$ & $420$ & $0.76$ &    $20.07_{-0.93}^{+0.93}$ &     $234.93_{-0.98}^{+0.94}$ &     $120$ & $-0.23$ & $-0.20$ \\
II.11 & $55764.2$ & $220$ & $0.56$ &    $6.02_{-5.54}^{+6.77}$ &     $315.53_{-4.10}^{+2.47}$ &     $10$ & $0.73$ & $-0.13$ \\
II.12 & $55911.3$ & $630$ & $0.84$ &    $18.49_{-1.43}^{+1.80}$ &     $36.56_{-1.54}^{+1.29}$ &     $-30$ & $-0.61$ & $-0.09$ \\
II.13 & $56146.2$ & $250$ & $0.59$ &    $1.60_{-0.25}^{+0.23}$ &     $329.68_{-0.31}^{+0.47}$ &     $-10$ & $0.03$ & $-0.42$ \\
II.14 & $56226.6$ & $700$ & $0.87$ &    $28.16_{-0.60}^{+0.43}$ &     $169.98_{-0.96}^{+0.71}$ &     $-50$ & $0.12$ & $0.37$ \\
II.15 & $56470.1$ & $660$ & $0.85$ &    $14.17_{-0.87}^{+1.08}$ &     $93.74_{-1.01}^{+0.83}$ &     $-10$ & $-0.29$ & $0.36$ \\
II.16 & $56521.8$ & $420$ & $0.76$ &    $-2.87_{-0.52}^{+0.46}$ &     $223.77_{-0.45}^{+0.42}$ &     $20$ & $1.57$ & $-1.12$ \\
II.17 & $56579.9$ & $210$ & $0.55$ &    $10.28_{-0.46}^{+0.35}$ &     $32.92_{-0.57}^{+0.74}$ &     $-180$ & $0.08$ & $-0.02$ \\
II.18 & $56666.5$ & $830$ & $0.88$ &    $33.02_{-0.43}^{+0.39}$ &     $293.12_{-0.99}^{+0.67}$ &     $-20$ & $0.20$ & $-0.17$ \\
II.19$^{ * }$ & $56757.1$ & $240$ & $0.59$ &        $81.22_{-5.86}^{+7.72}$ &         $2.11_{-47.87}^{+185.55}$ &         $+160$ & --- & --- \\
II.20 & $56800.0$ & $300$ & $0.64$ &    $17.90_{-1.12}^{+1.47}$ &     $349.50_{-2.69}^{+2.69}$ &     $-100$ & $-0.15$ & $0.11$ \\
II.21 & $56817.6$ & $340$ & $0.70$ &    $1.31_{-0.73}^{+0.84}$ &     $106.26_{-1.72}^{+2.20}$ &     $0$ & $0.02$ & $-0.00$ \\
II.22 & $56819.2$ & $4400$ & $0.99$ &    $11.45_{-0.15}^{+0.18}$ &     $110.65_{-0.58}^{+0.46}$ &     $-100$ & $0.03$ & $0.02$ \\
II.23 & $57157.9$ & $230$ & $0.59$ &    $12.14_{-0.43}^{+0.45}$ &     $91.49_{-0.63}^{+0.80}$ &     $-10$ & $-0.04$ & $-0.11$ \\
II.24 & $57217.9$ & $300$ & $0.63$ &    $26.36_{-1.81}^{+1.58}$ &     $326.29_{-1.08}^{+1.23}$ &     $0$ & $0.26$ & $0.79$ \\
II.25 & $57246.8$ & $370$ & $0.73$ &    $6.17_{-0.45}^{+0.42}$ &     $328.27_{-0.75}^{+0.61}$ &     $-10$ & $0.17$ & $-0.13$ \\
II.26 & $57269.8$ & $220$ & $0.57$ &    $28.08_{-0.42}^{+0.45}$ &     $133.77_{-0.71}^{+0.42}$ &     $0$ & $0.08$ & $-0.23$ \\
II.27 & $57312.7$ & $220$ & $0.58$ &    $19.95_{-1.94}^{+2.32}$ &     $197.53_{-2.08}^{+2.05}$ &     $-10$ & $0.05$ & $-0.07$ \\
II.28 & $57340.9$ & $730$ & $0.87$ &    $12.71_{-0.62}^{+0.56}$ &     $76.16_{-1.11}^{+1.15}$ &     $-10$ & $0.11$ & $-0.14$ \\
II.29 & $57478.6$ & $370$ & $0.73$ &    $15.48_{-0.64}^{+0.55}$ &     $151.22_{-0.53}^{+0.47}$ &     $-10$ & $-0.12$ & $-0.11$ \\
II.30 & $57672.1$ & $380$ & $0.74$ &    $1.16_{-2.60}^{+5.37}$ &     $32.08_{-4.41}^{+4.17}$ &     $50$ & $-25.44^{\ominus}$ & $22.38^{\ominus}$ \\
II.31$^{ ** }$ & $57951.8$ & $440$ & $0.77$ &    $25.16_{-1.15}^{+1.02}$ &     $208.39_{-0.98}^{+1.43}$ &     --- & --- & --- \\
II.32$^{ ** }$ & $58063.8$ & $1200$ & $0.93$ &    $7.44_{-0.24}^{+0.27}$ &     $340.14_{-0.53}^{+0.52}$ &     --- & --- & --- \\
II.33$^{ ** }$ & $58141.7$ & $290$ & $0.62$ &    $8.61_{-0.39}^{+0.57}$ &     $76.33_{-1.89}^{+1.88}$ &     --- & --- & --- \\
II.34$^{ ** }$ & $58205.1$ & $550$ & $0.82$ &    $7.44_{-1.27}^{+1.37}$ &     $307.09_{-4.00}^{+2.22}$ &     --- & --- & --- \\
II.35$^{ ** }$ & $58264.3$ & $390$ & $0.74$ &    $8.27_{-4.79}^{+1.49}$ &     $210.67_{-6.35}^{+5.56}$ &     --- & --- & --- \\
\hline
\end{longtable}
\caption{Table of observed events with $E_{\mu , \mathrm{proxy}} >   \SI{200}{\tera\eV}$, i.e. with a signalness  $S(E_{\mu}) = \frac{\Phi_\mathrm{signal}(E_{\mu})}{\Phi_\mathrm{signal}(E_{\mu}) + \Phi_\mathrm{background}(E_{\mu})} \gtrsim 0.5$. The reconstructed muon energy is as used in the analysis~\citep{abbasiImprovedMethodMeasuring2013}. The directional reconstruction is based on a more sophisticated reconstruction algorithm that is also applied to realtime alerts~\citep{aartsenIceCubeRealtimeAlert2017,icecubeMultimessengerObservationsFlaring2018}. The given statistical uncertainty ranges are $90\%$ CL, derived from reconstructions performed on a sample of similar events \citep{LeifObservationNuMu2017}. See text for a description of the Pass-2 re-calibration campaign that leads to changes of reconstructed energy and direction for some events. Seven events that have been reported in the last publications~\citep{aartsenObservationCharacterizationCosmic2016,haackMeasurementDiffuseAstrophysical2018} are marked as dropped, either because their reconstructed energy falls below threshold or because they do not pass the event selection anymore.\\ \label{tab:all_highe_events}\begin{minipage}{0.45\textwidth}
$^{\dagger}$Dropped: Pass-2 Energy below threshold \\
$^{\ddagger}$Dropped: Did not pass unified event selection \\
$^{*}$ New: Pass-2 Energy above threshold
\end{minipage} \hfill
\begin{minipage}{0.45\textwidth}
$^{\ominus}$ Preliminary direction was reported in~\cite{aartsenObservationCharacterizationCosmic2016}.\\
$^{**}$ New Event
\end{minipage}
}
\end{table*}

\end{appendices}
\FloatBarrier
\bibliography{References}   

\begin{thebibliography}{}
\expandafter\ifx\csname natexlab\endcsname\relax\def\natexlab#1{#1}\fi
\providecommand{\url}[1]{\href{#1}{#1}}
\providecommand{\dodoi}[1]{doi:~\href{http://doi.org/#1}{\nolinkurl{#1}}}
\providecommand{\doeprint}[1]{\href{http://ascl.net/#1}{\nolinkurl{http://ascl.net/#1}}}
\providecommand{\doarXiv}[1]{\href{https://arxiv.org/abs/#1}{\nolinkurl{https://arxiv.org/abs/#1}}}

\bibitem[{Aartsen {et~al.}(2021{\natexlab{a}})Aartsen, Abbasi, Ackermann,
  {et~al.}}]{AartsenGlashowDetectionIcecube2021}
Aartsen, M., Abbasi, R., Ackermann, M., {et~al.} 2021{\natexlab{a}}, Nature,
  591, 220–224, \dodoi{10.1038/s41586-021-03256-1}

\bibitem[{Aartsen {et~al.}(2021{\natexlab{b}})Aartsen, Abbasi, Ackermann,
  {et~al.}}]{AartsenGen2:2020}
---. 2021{\natexlab{b}}, J. Phys. G, 48, 060501,
  \dodoi{10.1088/1361-6471/abbd48}

\bibitem[{Aartsen {et~al.}(2016)Aartsen, Abraham, Ackermann,
  {et~al.}}]{aartsenObservationCharacterizationCosmic2016}
Aartsen, M., Abraham, K., Ackermann, M., {et~al.} 2016, The Astrophysical
  Journal, 833, 3, \dodoi{10.3847/0004-637X/833/1/3}

\bibitem[{Aartsen {et~al.}(2017{\natexlab{a}})Aartsen, Ackermann, Adams,
  {et~al.}}]{aartsenIceCubeNeutrinoObservatory2017}
Aartsen, M., Ackermann, M., Adams, J., {et~al.} 2017{\natexlab{a}}, J. Inst.,
  12, P03012, \dodoi{10.1088/1748-0221/12/03/P03012}

\bibitem[{Aartsen {et~al.}(2017{\natexlab{b}})Aartsen, Ackermann, Adams,
  {et~al.}}]{gcn17569642}
---. 2017{\natexlab{b}}, {GCN/AMON Notice 17569642\_130214 (EHE)},
  \url{https://gcn.gsfc.nasa.gov/notices_amon/17569642_130214.amon}

\bibitem[{Aartsen {et~al.}(2017{\natexlab{c}})Aartsen, Ackermann, Adams,
  {et~al.}}]{aartsenIceCubeRealtimeAlert2017}
---. 2017{\natexlab{c}}, Astroparticle Physics, 92, 30,
  \dodoi{10.1016/j.astropartphys.2017.05.002}

\bibitem[{Aartsen {et~al.}(2018{\natexlab{a}})Aartsen, Ackermann, Adams,
  {et~al.}}]{aartsenNeutrinoEmissionDirection2018}
---. 2018{\natexlab{a}}, Science, 361, 147, \dodoi{10.1126/science.aat2890}

\bibitem[{Aartsen {et~al.}(2020{\natexlab{a}})Aartsen, Ackermann, Adams,
  {et~al.}}]{Aartsen:2019fau}
---. 2020{\natexlab{a}}, Phys. Rev. Lett., 124, 051103,
  \dodoi{10.1103/PhysRevLett.124.051103}

\bibitem[{Aartsen {et~al.}(2020{\natexlab{b}})Aartsen, Ackermann, Adams,
  {et~al.}}]{aartsenInSituCalibration2020}
---. 2020{\natexlab{b}}, Journal of Instrumentation, 15, P06032,
  \dodoi{10.1088/1748-0221/15/06/p06032}

\bibitem[{Aartsen {et~al.}(2013)Aartsen, Abbasi, Abdou,
  {et~al.}}]{PhysRevLett.111.021103}
Aartsen, M.~G., Abbasi, R., Abdou, Y., {et~al.} 2013, Phys. Rev. Lett., 111,
  021103, \dodoi{10.1103/PhysRevLett.111.021103}

\bibitem[{Aartsen {et~al.}(2017{\natexlab{d}})Aartsen, Ackermann, Adams,
  {et~al.}}]{aartsenGRBlimit2017}
Aartsen, M.~G., Ackermann, M., Adams, J., {et~al.} 2017{\natexlab{d}},
  Astrophys. J., 843, 112, \dodoi{10.3847/1538-4357/aa7569}

\bibitem[{Aartsen {et~al.}(2018{\natexlab{b}})Aartsen, Ackermann, Adams,
  {et~al.}}]{icecubeMultimessengerObservationsFlaring2018}
---. 2018{\natexlab{b}}, Science, 361, \dodoi{10.1126/science.aat1378}

\bibitem[{Aartsen {et~al.}(2019)Aartsen, Ackermann, Adams,
  {et~al.}}]{Aartsen:2018vezInelasticity}
---. 2019, Phys. Rev. D, 99, 032004, \dodoi{10.1103/PhysRevD.99.032004}

\bibitem[{Aartsen {et~al.}(2020{\natexlab{c}})Aartsen, Ackermann, Adams,
  {et~al.}}]{aartsenCharacteristicsDiffuseAstrophysical2020}
---. 2020{\natexlab{c}}, Phys. Rev. Lett., 125, 121104,
  \dodoi{10.1103/PhysRevLett.125.121104}

\bibitem[{Abbasi {et~al.}(2013)Abbasi, Abdou,
  {et~al.}}]{abbasiImprovedMethodMeasuring2013}
Abbasi, R., Abdou, Y.and~Ackermann, M., {et~al.} 2013, Nuclear Instruments and
  Methods in Physics Research Section A, 703, 190,
  \dodoi{10.1016/j.nima.2012.11.081}

\bibitem[{Abbasi {et~al.}(2010)Abbasi, Abdou, Abu-Zayyad,
  {et~al.}}]{Abbasi:2010PMT}
Abbasi, R., Abdou, Y., Abu-Zayyad, T., {et~al.} 2010, Nuclear Instruments and
  Methods in Physics Research Section A, 618, 139,
  \dodoi{https://doi.org/10.1016/j.nima.2010.03.102}

\bibitem[{Abbasi {et~al.}(2012{\natexlab{a}})Abbasi, Abdou, Abu-Zayyad,
  {et~al.}}]{AbbasiAbsenceNuInGRB2012Nature}
---. 2012{\natexlab{a}}, \nat, 484, 351, \dodoi{10.1038/nature11068}

\bibitem[{Abbasi {et~al.}(2012{\natexlab{b}})Abbasi, Abdou, Abu-Zayyad,
  {et~al.}}]{AbbasiDeepCore2011}
---. 2012{\natexlab{b}}, Astropart. Phys., 35, 615,
  \dodoi{10.1016/j.astropartphys.2012.01.004}

\bibitem[{Abbasi {et~al.}(2009)Abbasi, Ackermann, Adams,
  {et~al.}}]{abbasiIceCubeDataAcquisition2009}
Abbasi, R., Ackermann, M., Adams, J., {et~al.} 2009, Nuclear Instruments and
  Methods in Physics Research Section A, 601, 294,
  \dodoi{10.1016/j.nima.2009.01.001}

\bibitem[{Abbasi {et~al.}(2021)Abbasi, Ackermann, Adams,
  {et~al.}}]{Abbasi:2020jmhHESE75}
---. 2021, Phys. Rev. D, 104, 022002, \dodoi{10.1103/PhysRevD.104.022002}

\bibitem[{{Ahlers} \& {Halzen}(2018)}]{HalzenOpeningWindowUniverseIceCube2018}
{Ahlers}, M., \& {Halzen}, F. 2018, Progress in Particle and Nuclear Physics,
  102, 73, \dodoi{10.1016/j.ppnp.2018.05.001}

\bibitem[{Ahrens {et~al.}(2004)Ahrens, Bai, Bay,
  {et~al.}}]{ahrensMuonTrackReconstruction2004}
Ahrens, J., Bai, X., Bay, R., {et~al.} 2004, Nuclear Instruments and Methods in
  Physics Research Section A, 524, 169, \dodoi{10.1016/j.nima.2004.01.065}

\bibitem[{Barr {et~al.}(2006)Barr, Robbins, Gaisser, \&
  Stanev}]{barrUncertaintiesAtmosphericNeutrino2006}
Barr, G.~D., Robbins, S., Gaisser, T.~K., \& Stanev, T. 2006, Phys. Rev. D, 74,
  094009, \dodoi{10.1103/PhysRevD.74.094009}

\bibitem[{{Becker}(2008)}]{BeckerHighEnergyNuMultimessenenger2008}
{Becker}, J.~K. 2008, \physrep, 458, 173, \dodoi{10.1016/j.physrep.2007.10.006}

\bibitem[{{Becker Tjus} \&
  {Merten}(2020)}]{BeckerClosingInOnOriginGalacticCR2020}
{Becker Tjus}, J., \& {Merten}, L. 2020, \physrep, 872, 1,
  \dodoi{10.1016/j.physrep.2020.05.002}

\bibitem[{Bhattacharya {et~al.}(2015)Bhattacharya, Enberg, Reno, Sarcevic, \&
  Stasto}]{bhattacharyaPerturbativeCharmProduction2015}
Bhattacharya, A., Enberg, R., Reno, M.~H., Sarcevic, I., \& Stasto, A. 2015, J.
  High Energ. Phys., 2015, 110, \dodoi{10.1007/JHEP06(2015)110}

\bibitem[{Biehl {et~al.}(2018)Biehl, Boncioli, Fedynitch, \&
  Winter}]{biehlCosmicRayNeutrinoEmission2018}
Biehl, D., Boncioli, D., Fedynitch, A., \& Winter, W. 2018, A\&A, 611, A101,
  \dodoi{10.1051/0004-6361/201731337}

\bibitem[{{Biehl} {et~al.}(2018){Biehl}, {Boncioli}, {Lunardini}, \&
  {Winter}}]{BiehlTDStarsAsCRandNuOrigin2018}
{Biehl}, D., {Boncioli}, D., {Lunardini}, C., \& {Winter}, W. 2018, Scientific
  Reports, 8, 10828, \dodoi{10.1038/s41598-018-29022-4}

\bibitem[{Chirkin(2015)}]{Chirkin:2015kga}
Chirkin, D. 2015, in {GPU Computing in High-Energy Physics},
  \dodoi{10.3204/DESY-PROC-2014-05/40}

\bibitem[{{Cooper-Sarkar} {et~al.}(2011){Cooper-Sarkar}, Mertsch, \&
  Sarkar}]{coopersarkarHighEnergyNeutrino2011}
{Cooper-Sarkar}, A., Mertsch, P., \& Sarkar, S. 2011, J. High Energ. Phys.,
  2011, 42, \dodoi{10.1007/JHEP08(2011)042}

\bibitem[{{Dai} \& {Fang}(2017)}]{DaiTDEproduceIceCubeNu2017}
{Dai}, L., \& {Fang}, K. 2017, \mnras, 469, 1354, \dodoi{10.1093/mnras/stx863}

\bibitem[{Eller(2019)}]{ellerUnifiedHoleiceModel2019}
Eller, P. 2019, {Unified Hole-ice Model: angular-acceptance code},
  https://github.com/philippeller/angular\_acceptance

\bibitem[{Enberg {et~al.}(2008)Enberg, Reno, \&
  Sarcevic}]{enbergPromptNeutrinoFluxes2008}
Enberg, R., Reno, M.~H., \& Sarcevic, I. 2008, Phys. Rev. D, 78, 043005,
  \dodoi{10.1103/PhysRevD.78.043005}

\bibitem[{{Farrar} \& {Piran}(2014)}]{FarrarTDEJetsUHECRSource2014}
{Farrar}, G.~R., \& {Piran}, T. 2014, arXiv e-prints, arXiv:1411.0704.
\newblock \doarXiv{1411.0704}

\bibitem[{Fedynitch {et~al.}(2015)Fedynitch, Engel, Gaisser, Riehn, \&
  Stanev}]{fedynitchCalculationConventionalPrompt2015}
Fedynitch, A., Engel, R., Gaisser, T.~K., Riehn, F., \& Stanev, T. 2015, EPJ
  Web Conf., 99, 08001, \dodoi{10.1051/epjconf/20159908001}

\bibitem[{Fedynitch {et~al.}(2019)Fedynitch, Riehn, Engel, Gaisser, \&
  Stanev}]{fedynitchHadronicInteractionModel2019}
Fedynitch, A., Riehn, F., Engel, R., Gaisser, T.~K., \& Stanev, T. 2019, Phys.
  Rev. D, 100, 103018, \dodoi{10.1103/PhysRevD.100.103018}

\bibitem[{Fusco \& Versari(2019)}]{fuscoStudyHighenergyNeutrino2019}
Fusco, L.~A., \& Versari, F. 2019, in {Proceedings of Science},
  \dodoi{10.22323/1.358.0891}

\bibitem[{Gaisser(2012)}]{gaisserSpectrumCosmicrayNucleons2012}
Gaisser, T.~K. 2012, Astroparticle Physics, 35, 801,
  \dodoi{10.1016/j.astropartphys.2012.02.010}

\bibitem[{Gaisser {et~al.}(2013)Gaisser, Stanev, \&
  Tilav}]{gaisserCosmicRayEnergy2013}
Gaisser, T.~K., Stanev, T., \& Tilav, S. 2013, Frontiers of Physics,
  \dodoi{10.1007/s11467-013-0319-7}

\bibitem[{{Gu{\'e}pin} \&
  {Kotera}(2017)}]{GuepinObserveNuFlaresCoincWExplosiveTransients2017}
{Gu{\'e}pin}, C., \& {Kotera}, K. 2017, \aap, 603, A76,
  \dodoi{10.1051/0004-6361/201630326}

\bibitem[{Haack \& others
  (IceCube~Collaboration)(2018)}]{haackMeasurementDiffuseAstrophysical2018}
Haack, C., \& others (IceCube~Collaboration). 2018, in {Proceedings of 35th
  International Cosmic Ray Conference \textemdash{} PoS(ICRC2017)}, Vol. 301
  ({SISSA Medialab}), 1005, \dodoi{10.22323/1.301.1005}

\bibitem[{{Halzen} \& {Klein}(2010)}]{HalzenIceCubeInstrumentNuAstronomy2010}
{Halzen}, F., \& {Klein}, S.~R. 2010, Review of Scientific Instruments, 81,
  081101, \dodoi{10.1063/1.3480478}

\bibitem[{Hayasaki(2021)}]{HayasakiNufromTDE2021}
Hayasaki, K. 2021, Nature Astronomy, 5, 436, \dodoi{10.1038/s41550-021-01309-z}

\bibitem[{{Heck} {et~al.}(1998){Heck}, Knapp, Capdevielle,
  {et~al.}}]{corsika1998}
{Heck}, D., Knapp, J., Capdevielle, J., {et~al.} 1998, Tech. Rep. FZKA, 6019

\bibitem[{Honda {et~al.}(2007)Honda, Kajita, Kasahara, Midorikawa, \&
  Sanuki}]{Honda:Atmosflux2007}
Honda, M., Kajita, T., Kasahara, K., Midorikawa, S., \& Sanuki, T. 2007, Phys.
  Rev. D, 75, 043006, \dodoi{10.1103/PhysRevD.75.043006}

\bibitem[{Kimura {et~al.}(2015)Kimura, Murase, \&
  Toma}]{kimuraNeutrinoCosmicrayEmission2015}
Kimura, S.~S., Murase, K., \& Toma, K. 2015, ApJ, 806, 159,
  \dodoi{10.1088/0004-637X/806/2/159}

\bibitem[{Kopper(2019)}]{kopperCLSimCode2019}
Kopper, C. 2019, {{CLSim}} - {{Code}}, https://github.com/claudiok/clsim

\bibitem[{Köhne {et~al.}(2013)Köhne, Frantzen, Schmitz, Fuchs, Rhode,
  Chirkin, \& Becker~Tjus}]{koehnePROPOSALToolPropagation2013}
Köhne, J.~H., Frantzen, K., Schmitz, M., {et~al.} 2013, Computer Physics
  Communications, 184, 2070, \dodoi{10.1016/j.cpc.2013.04.001}

\bibitem[{Learned \&
  Mannheim(2000)}]{learnedHighEnergyNeutrinoAstrophysics2000}
Learned, J.~G., \& Mannheim, K. 2000, Annu. Rev. Nucl. Part. Sci., 50, 679,
  \dodoi{10.1146/annurev.nucl.50.1.679}

\bibitem[{Liu {et~al.}(2018)Liu, Murase, Inoue, Ge, \&
  Wang}]{liuCanWindsDriven2018}
Liu, R.-Y., Murase, K., Inoue, S., Ge, C., \& Wang, X.-Y. 2018, ApJ, 858, 9,
  \dodoi{10.3847/1538-4357/aaba74}

\bibitem[{{Loeb} \& {Waxman}(2006)}]{LoebBackgroundNuFromStarburstGalaxies2006}
{Loeb}, A., \& {Waxman}, E. 2006, \jcap, 2006, 003,
  \dodoi{10.1088/1475-7516/2006/05/003}

\bibitem[{{Lunardini} \& {Winter}(2017)}]{LunardiniHENuTDEStars2017}
{Lunardini}, C., \& {Winter}, W. 2017, \prd, 95, 123001,
  \dodoi{10.1103/PhysRevD.95.123001}

\bibitem[{Murase {et~al.}(2014)Murase, Inoue, \&
  Dermer}]{muraseDiffuseNeutrinoIntensity2014}
Murase, K., Inoue, Y., \& Dermer, C.~D. 2014, Phys. Rev. D, 90, 023007,
  \dodoi{10.1103/PhysRevD.90.023007}

\bibitem[{Padovani {et~al.}(2015)Padovani, Petropoulou, Giommi, \&
  Resconi}]{padovaniSimplifiedViewBlazars2015}
Padovani, P., Petropoulou, M., Giommi, P., \& Resconi, E. 2015, Monthly Notices
  of the Royal Astronomical Society, 452, 1877, \dodoi{10.1093/mnras/stv1467}

\bibitem[{Rongen(2019)}]{rongenCalibrationIceCubeNeutrino2019}
Rongen, M. 2019, Dissertation, RWTH Aachen University,
  \dodoi{10.18154/RWTH-2019-09941}

\bibitem[{Rädel(2017)}]{LeifObservationNuMu2017}
Rädel, L. 2017, Dissertation, RWTH Aachen University,
  \dodoi{10.18154/RWTH-2017-10054}

\bibitem[{Schatto(2014)}]{kaiStackedSearchesHighenergy2014}
Schatto, K. 2014, Dissertation, University Mainz

\bibitem[{Senno {et~al.}(2015)Senno, M{\'e}sz{\'a}ros, Murase, Baerwald, \&
  Rees}]{sennoExtragalacticStarformingGalaxies2015}
Senno, N., M{\'e}sz{\'a}ros, P., Murase, K., Baerwald, P., \& Rees, M.~J. 2015,
  The Astrophysical Journal, 806, 24, \dodoi{10.1088/0004-637X/806/1/24}

\bibitem[{Senno {et~al.}(2016)Senno, Murase, \&
  Meszaros}]{sennoChokedJetsLowLuminosity2016}
Senno, N., Murase, K., \& Meszaros, P. 2016, Physical Review D, 93,
  \dodoi{10.1103/PhysRevD.93.083003}

\bibitem[{{Senno} {et~al.}(2017){Senno}, {Murase}, \&
  {M{\'e}sz{\'a}ros}}]{SennoNuFlaresXRayTDEs2017}
{Senno}, N., {Murase}, K., \& {M{\'e}sz{\'a}ros}, P. 2017, \apj, 838, 3,
  \dodoi{10.3847/1538-4357/aa6344}

\bibitem[{{Stein} {et~al.}(2021){Stein}, {Velzen}, {Kowalski}, {Franckowiak},
  {Gezari}, {Miller-Jones}, {Frederick}, {Sfaradi}, {Bietenholz}, {Horesh},
  {Fender}, {Garrappa}, {Ahumada}, {Andreoni}, {Belicki}, {Bellm},
  {B{\"o}ttcher}, {Brinnel}, {Burruss}, {Cenko}, {Coughlin}, {Cunningham},
  {Drake}, {Farrar}, {Feeney}, {Foley}, {Gal-Yam}, {Golkhou}, {Goobar},
  {Graham}, {Hammerstein}, {Helou}, {Hung}, {Kasliwal}, {Kilpatrick}, {Kong},
  {Kupfer}, {Laher}, {Mahabal}, {Masci}, {Necker}, {Nordin}, {Perley},
  {Rigault}, {Reusch}, {Rodriguez}, {Rojas-Bravo}, {Rusholme}, {Shupe},
  {Singer}, {Sollerman}, {Soumagnac}, {Stern}, {Taggart}, {van Santen}, {Ward},
  {Woudt}, \& {Yao}}]{SteinTDECoincHENu2021}
{Stein}, R., {Velzen}, S.~v., {Kowalski}, M., {et~al.} 2021, Nature Astronomy,
  5, 510, \dodoi{10.1038/s41550-020-01295-8}

\bibitem[{Stettner \& others
  (IceCube~Collaboration)(2019)}]{stettner2019measurement}
Stettner, J., \& others (IceCube~Collaboration). 2019, Proceedings of Science,
  ICRC-2019.
\newblock \doarXiv{1908.09551}

\bibitem[{Stettner(2021)}]{stettner2021thesis}
Stettner, J.~B. 2021, Dissertation, RWTH Aachen University, Aachen,
  \dodoi{10.18154/RWTH-2021-01139}

\bibitem[{{Tamborra} {et~al.}(2014){Tamborra}, {Ando}, \&
  {Murase}}]{KohtaStarFormingGalaxiesAsDiffuse2014}
{Tamborra}, I., {Ando}, S., \& {Murase}, K. 2014, \jcap, 2014, 043,
  \dodoi{10.1088/1475-7516/2014/09/043}

\bibitem[{Tavecchio \&
  Ghisellini(2014)}]{tavecchioHighenergyCosmicNeutrinos2014}
Tavecchio, F., \& Ghisellini, G. 2014, Monthly Notices of the Royal
  Astronomical Society

\bibitem[{Wilks(1938)}]{wilksLargeSampleDistributionLikelihood1938}
Wilks, S.~S. 1938, Ann. Math. Statist., 9, 60, \dodoi{10.1214/aoms/1177732360}

\end{thebibliography}

\end{document}